\documentstyle[10pt,epsf,epsfig,hangcaption,xspace,amssymb,amsfonts,amsmath,amsthm,cite,
dp_delphititle,lineno]{dp_delphi}
\makeindex
\def\DpPaperGroup{EP}
\def\DpPaperRef{2003-064}
\def\DpDate{23 September 2003}
\def\DpAuthors{DELPHI Collaboration}
\def\DpSubmit{(Accepted by Eur. Phys. J. C )}
\def\DpTitle{{ Search for Charged Higgs Bosons at LEP in General Two Higgs
Doublet Models}}
\def\DpComment{ }
\def\DpEMail{ }

\newcommand{\hphm}{\mathrm{H}^+\mathrm{H}^-}
\newcommand{\ww}{\mathrm{W}^+\mathrm{W}^-}
\newcommand{\zo}{\mathrm{Z}^0}
\newcommand{\zz}{\zo\zo}
\newcommand{\cs}{\bar{\mathrm{c}}\mathrm{s}}
\newcommand{\csb}{\mathrm{c}\bar{\mathrm{s}}}
\newcommand{\tn}{\tau^- \bar{\nu}_{\tau}}
\newcommand{\tnp}{\tau^+ \nu_{\tau}}
\newcommand{\tto}{\tau^+ \tau^-}
\newcommand{\ee}{\mathrm{e}^- \mathrm{e}^+}
\newcommand{\HTT}{\tnp \tn}
\newcommand{\HCSTN}{\csb\tn}
\newcommand{\HCSCS}{\csb \cs}
\newcommand{\HWAWA}{\mathrm{W}^*\!\mathrm{A}\mathrm{W}^*\!\mathrm{A}}
\newcommand{\wa}{\mathrm{W}^*\!A}
\newcommand{\HWATN}{\mathrm{W}^*\!\mathrm{A}\tn}
\newcommand{\HWACS}{\mathrm{W}^*\!\mathrm{A} \cs}
\newcommand{\qq}{\mathrm{q}\bar{\mathrm{q}}}
\newcommand{\qqg} {\mbox{$ {\mathrm q}\bar{\mathrm q}(\gamma) $}}
\newcommand{\bb}{\mathrm{b}\bar{\mathrm{b}}}
\newcommand{\sqs}{\sqrt{s}}
\newcommand{\m}{\rm M}
\newcommand{\ma}{\m_{\mathrm{A}}}
\newcommand{\mz}{\m_{\mathrm{Z}}}
\newcommand{\mhp}{\m_{{\rm H}}}
\newcommand{\gev}{{\ifmmode \mbox{Ge\kern-0.2exV}
\else Ge\kern-0.2exV\nolinebreak\fi}}
\newcommand{\mev}{{\ifmmode \mbox{Me\kern-0.2exV}
\else Me\kern-0.2exV\nolinebreak\fi}}

\begin{document}
\makeatletter
\newcount\@tempcntc
\def\@citex[#1]#2{\if@filesw\immediate\write\@auxout{\string\citation{#2}}\fi
  \@tempcnta\z@\@tempcntb\m@ne\def\@citea{}\@cite{\@for\@citeb:=#2\do
    {\@ifundefined
       {b@\@citeb}{\@citeo\@tempcntb\m@ne\@citea\def\@citea{,}{\bf ?}\@warning
       {Citation `\@citeb' on page \thepage \space undefined}}%
    {\setbox\z@\hbox{\global\@tempcntc0\csname b@\@citeb\endcsname\relax}%
     \ifnum\@tempcntc=\z@ \@citeo\@tempcntb\m@ne
       \@citea\def\@citea{,}\hbox{\csname b@\@citeb\endcsname}%
     \else
      \advance\@tempcntb\@ne
      \ifnum\@tempcntb=\@tempcntc
      \else\advance\@tempcntb\m@ne\@citeo
      \@tempcnta\@tempcntc\@tempcntb\@tempcntc\fi\fi}}\@citeo}{#1}}
\def\@citeo{\ifnum\@tempcnta>\@tempcntb\else\@citea\def\@citea{,}%
  \ifnum\@tempcnta=\@tempcntb\the\@tempcnta\else
   {\advance\@tempcnta\@ne\ifnum\@tempcnta=\@tempcntb \else \def\@citea{--}\fi
    \advance\@tempcnta\m@ne\the\@tempcnta\@citea\the\@tempcntb}\fi\fi}
 
\makeatother
\begin{titlepage}
\pagenumbering{roman}
\CERNpreprint{\DpPaperGroup}{\DpPaperRef} 
\date{{\small\DpDate}} 
\title{\DpTitle} 
\address{\DpAuthors} 
\begin{shortabs} 
\noindent
%
\noindent

A search for pair-produced charged Higgs bosons was performed in the
data collected by the DELPHI detector at LEP II at
centre-of-mass energies from 189~GeV to 209~GeV\@. Five different
final states, $\HTT$, $\HCSCS$, $\HCSTN$, $\HWAWA$ and $\HWATN$ were considered,
 accounting for the major expected 
decays in type I and type II Two Higgs Doublet Models. 
No significant
excess of data compared to the expected Standard Model processes was
observed. The existence of a charged Higgs boson with mass lower
than 76.7 GeV/$c^2$ (type I) or 74.4 GeV/$c^2$ (type II) is excluded at the 
95\% confidence level, for a wide range of the model parameters.
Model independent cross-section limits have also been calculated.
\end{shortabs}
\vfill
\begin{center}
\DpSubmit \ \\ 
\DpComment \ \\
\DpEMail \ \\
\end{center}
\vfill
\clearpage
\headsep 10.0pt
\addtolength{\textheight}{10mm}
\addtolength{\footskip}{-5mm}
\begingroup
%
\newcommand{\DpName}[2]{\hbox{#1$^{\ref{#2}}$},\hfill}
\newcommand{\DpNameTwo}[3]{\hbox{#1$^{\ref{#2},\ref{#3}}$},\hfill}
\newcommand{\DpNameThree}[4]{\hbox{#1$^{\ref{#2},\ref{#3},\ref{#4}}$},\hfill}
\newskip\Bigfill \Bigfill = 0pt plus 1000fill
\newcommand{\DpNameLast}[2]{\hbox{#1$^{\ref{#2}}$}\hspace{\Bigfill}}
%
\footnotesize
\noindent
\DpName{J.Abdallah}{LPNHE}
\DpName{P.Abreu}{LIP}
\DpName{W.Adam}{VIENNA}
\DpName{P.Adzic}{DEMOKRITOS}
\DpName{T.Albrecht}{KARLSRUHE}
\DpName{T.Alderweireld}{AIM}
\DpName{R.Alemany-Fernandez}{CERN}
\DpName{T.Allmendinger}{KARLSRUHE}
\DpName{P.P.Allport}{LIVERPOOL}
\DpName{U.Amaldi}{MILANO2}
% ...
\normalsize
\endgroup
\titlefoot{Department of Physics and Astronomy, Iowa State
     University, Ames IA 50011-3160, USA
    \label{AMES}}
\titlefoot{Physics Department, Universiteit Antwerpen,
     Universiteitsplein 1, B-2610 Antwerpen, Belgium \\
     \indent~~and IIHE, ULB-VUB,
     Pleinlaan 2, B-1050 Brussels, Belgium \\
     \indent~~and Facult\'e des Sciences,
     Univ. de l'Etat Mons, Av. Maistriau 19, B-7000 Mons, Belgium
    \label{AIM}}
\titlefoot{Physics Laboratory, University of Athens, Solonos Str.
     104, GR-10680 Athens, Greece
    \label{ATHENS}}
\titlefoot{Department of Physics, University of Bergen,
     All\'egaten 55, NO-5007 Bergen, Norway
    \label{BERGEN}}
\titlefoot{Dipartimento di Fisica, Universit\`a di Bologna and INFN,
     Via Irnerio 46, IT-40126 Bologna, Italy
    \label{BOLOGNA}}
\titlefoot{Centro Brasileiro de Pesquisas F\'{\i}sicas, rua Xavier Sigaud 150,
     BR-22290 Rio de Janeiro, Brazil \\
     \indent~~and Depto. de F\'{\i}sica, Pont. Univ. Cat\'olica,
     C.P. 38071 BR-22453 Rio de Janeiro, Brazil \\
     \indent~~and Inst. de F\'{\i}sica, Univ. Estadual do Rio de Janeiro,
     rua S\~{a}o Francisco Xavier 524, Rio de Janeiro, Brazil
    \label{BRASIL}}
\titlefoot{Coll\`ege de France, Lab. de Physique Corpusculaire, IN2P3-CNRS,
     FR-75231 Paris Cedex 05, France
    \label{CDF}}
\titlefoot{CERN, CH-1211 Geneva 23, Switzerland
    \label{CERN}}
\titlefoot{Institut de Recherches Subatomiques, IN2P3 - CNRS/ULP - BP20,
     FR-67037 Strasbourg Cedex, France
    \label{CRN}}
\titlefoot{Now at DESY-Zeuthen, Platanenallee 6, D-15735 Zeuthen, Germany
    \label{DESY}}
\titlefoot{Institute of Nuclear Physics, N.C.S.R. Demokritos,
     P.O. Box 60228, GR-15310 Athens, Greece
    \label{DEMOKRITOS}}
\titlefoot{FZU, Inst. of Phys. of the C.A.S. High Energy Physics Division,
     Na Slovance 2, CZ-180 40, Praha 8, Czech Republic
    \label{FZU}}
\titlefoot{Dipartimento di Fisica, Universit\`a di Genova and INFN,
     Via Dodecaneso 33, IT-16146 Genova, Italy
    \label{GENOVA}}
\titlefoot{Institut des Sciences Nucl\'eaires, IN2P3-CNRS, Universit\'e
     de Grenoble 1, FR-38026 Grenoble Cedex, France
    \label{GRENOBLE}}
\titlefoot{Helsinki Institute of Physics, P.O. Box 64,
     FIN-00014 University of Helsinki, Finland
    \label{HELSINKI}}
\titlefoot{Joint Institute for Nuclear Research, Dubna, Head Post
     Office, P.O. Box 79, RU-101 000 Moscow, Russian Federation
    \label{JINR}}
\titlefoot{Institut f\"ur Experimentelle Kernphysik,
     Universit\"at Karlsruhe, Postfach 6980, DE-76128 Karlsruhe,
     Germany
    \label{KARLSRUHE}}
\titlefoot{Institute of Nuclear Physics,Ul. Kawiory 26a,
     PL-30055 Krakow, Poland
    \label{KRAKOW1}}
\titlefoot{Faculty of Physics and Nuclear Techniques, University of Mining
     and Metallurgy, PL-30055 Krakow, Poland
    \label{KRAKOW2}}
\titlefoot{Universit\'e de Paris-Sud, Lab. de l'Acc\'el\'erateur
     Lin\'eaire, IN2P3-CNRS, B\^{a}t. 200, FR-91405 Orsay Cedex, France
    \label{LAL}}
\titlefoot{School of Physics and Chemistry, University of Lancaster,
     Lancaster LA1 4YB, UK
    \label{LANCASTER}}
\titlefoot{LIP, IST, FCUL - Av. Elias Garcia, 14-$1^{o}$,
     PT-1000 Lisboa Codex, Portugal
    \label{LIP}}
\titlefoot{Department of Physics, University of Liverpool, P.O.
     Box 147, Liverpool L69 3BX, UK
    \label{LIVERPOOL}}
\titlefoot{Dept. of Physics and Astronomy, Kelvin Building,
     University of Glasgow, Glasgow G12 8QQ
    \label{GLASGOW}}
\titlefoot{LPNHE, IN2P3-CNRS, Univ.~Paris VI et VII, Tour 33 (RdC),
     4 place Jussieu, FR-75252 Paris Cedex 05, France
    \label{LPNHE}}
\titlefoot{Department of Physics, University of Lund,
     S\"olvegatan 14, SE-223 63 Lund, Sweden
    \label{LUND}}
\titlefoot{Universit\'e Claude Bernard de Lyon, IPNL, IN2P3-CNRS,
     FR-69622 Villeurbanne Cedex, France
    \label{LYON}}
\titlefoot{Dipartimento di Fisica, Universit\`a di Milano and INFN-MILANO,
     Via Celoria 16, IT-20133 Milan, Italy
    \label{MILANO}}
\titlefoot{Dipartimento di Fisica, Univ. di Milano-Bicocca and
     INFN-MILANO, Piazza della Scienza 2, IT-20126 Milan, Italy
    \label{MILANO2}}
\titlefoot{IPNP of MFF, Charles Univ., Areal MFF,
     V Holesovickach 2, CZ-180 00, Praha 8, Czech Republic
    \label{NC}}
\titlefoot{NIKHEF, Postbus 41882, NL-1009 DB
     Amsterdam, The Netherlands
    \label{NIKHEF}}
\titlefoot{National Technical University, Physics Department,
     Zografou Campus, GR-15773 Athens, Greece
    \label{NTU-ATHENS}}
\titlefoot{Physics Department, University of Oslo, Blindern,
     NO-0316 Oslo, Norway
    \label{OSLO}}
\titlefoot{Dpto. Fisica, Univ. Oviedo, Avda. Calvo Sotelo
     s/n, ES-33007 Oviedo, Spain
    \label{OVIEDO}}
\titlefoot{Department of Physics, University of Oxford,
     Keble Road, Oxford OX1 3RH, UK
    \label{OXFORD}}
\titlefoot{Dipartimento di Fisica, Universit\`a di Padova and
     INFN, Via Marzolo 8, IT-35131 Padua, Italy
    \label{PADOVA}}
\titlefoot{Rutherford Appleton Laboratory, Chilton, Didcot
     OX11 OQX, UK
    \label{RAL}}
\titlefoot{Dipartimento di Fisica, Universit\`a di Roma II and
     INFN, Tor Vergata, IT-00173 Rome, Italy
    \label{ROMA2}}
\titlefoot{Dipartimento di Fisica, Universit\`a di Roma III and
     INFN, Via della Vasca Navale 84, IT-00146 Rome, Italy
    \label{ROMA3}}
\titlefoot{DAPNIA/Service de Physique des Particules,
     CEA-Saclay, FR-91191 Gif-sur-Yvette Cedex, France
    \label{SACLAY}}
\titlefoot{Instituto de Fisica de Cantabria (CSIC-UC), Avda.
     los Castros s/n, ES-39006 Santander, Spain
    \label{SANTANDER}}
\titlefoot{Inst. for High Energy Physics, Serpukov
     P.O. Box 35, Protvino, (Moscow Region), Russian Federation
    \label{SERPUKHOV}}
\titlefoot{J. Stefan Institute, Jamova 39, SI-1000 Ljubljana, Slovenia
     and Laboratory for Astroparticle Physics,\\
     \indent~~Nova Gorica Polytechnic, Kostanjeviska 16a, SI-5000 Nova Gorica, Slovenia, \\
     \indent~~and Department of Physics, University of Ljubljana,
     SI-1000 Ljubljana, Slovenia
    \label{SLOVENIJA}}
\titlefoot{Fysikum, Stockholm University,
     Box 6730, SE-113 85 Stockholm, Sweden
    \label{STOCKHOLM}}
\titlefoot{Dipartimento di Fisica Sperimentale, Universit\`a di
     Torino and INFN, Via P. Giuria 1, IT-10125 Turin, Italy
    \label{TORINO}}
\titlefoot{INFN,Sezione di Torino, and Dipartimento di Fisica Teorica,
     Universit\`a di Torino, Via P. Giuria 1,\\
     \indent~~IT-10125 Turin, Italy
    \label{TORINOTH}}
\titlefoot{Dipartimento di Fisica, Universit\`a di Trieste and
     INFN, Via A. Valerio 2, IT-34127 Trieste, Italy \\
     \indent~~and Istituto di Fisica, Universit\`a di Udine,
     IT-33100 Udine, Italy
    \label{TU}}
\titlefoot{Univ. Federal do Rio de Janeiro, C.P. 68528
     Cidade Univ., Ilha do Fund\~ao
     BR-21945-970 Rio de Janeiro, Brazil
    \label{UFRJ}}
\titlefoot{Department of Radiation Sciences, University of
     Uppsala, P.O. Box 535, SE-751 21 Uppsala, Sweden
    \label{UPPSALA}}
\titlefoot{IFIC, Valencia-CSIC, and D.F.A.M.N., U. de Valencia,
     Avda. Dr. Moliner 50, ES-46100 Burjassot (Valencia), Spain
    \label{VALENCIA}}
\titlefoot{Institut f\"ur Hochenergiephysik, \"Osterr. Akad.
     d. Wissensch., Nikolsdorfergasse 18, AT-1050 Vienna, Austria
    \label{VIENNA}}
\titlefoot{Inst. Nuclear Studies and University of Warsaw, Ul.
     Hoza 69, PL-00681 Warsaw, Poland
    \label{WARSZAWA}}
\titlefoot{Fachbereich Physik, University of Wuppertal, Postfach
     100 127, DE-42097 Wuppertal, Germany
    \label{WUPPERTAL}}
\addtolength{\textheight}{-10mm}
\addtolength{\footskip}{5mm}
\clearpage
\headsep 30.0pt
\end{titlepage}
%
\pagenumbering{arabic} 
\setcounter{footnote}{0} %
\large
\section{Introduction}

A search for pair-produced charged Higgs bosons 
in e$^+$e$^-$ collisions
was performed 
using
the data collected by DELPHI during the LEP runs at centre-of-mass
energies from 189~GeV to 209~GeV\@. The results reported here supersede
those obtained in an earlier analysis of the DELPHI data~\cite{delphihh}.  
Similar searches have been
performed by the other LEP experiments~\cite{other_lep}.

\begin{figure}[hbpt]
\begin{center}
\epsfig{file=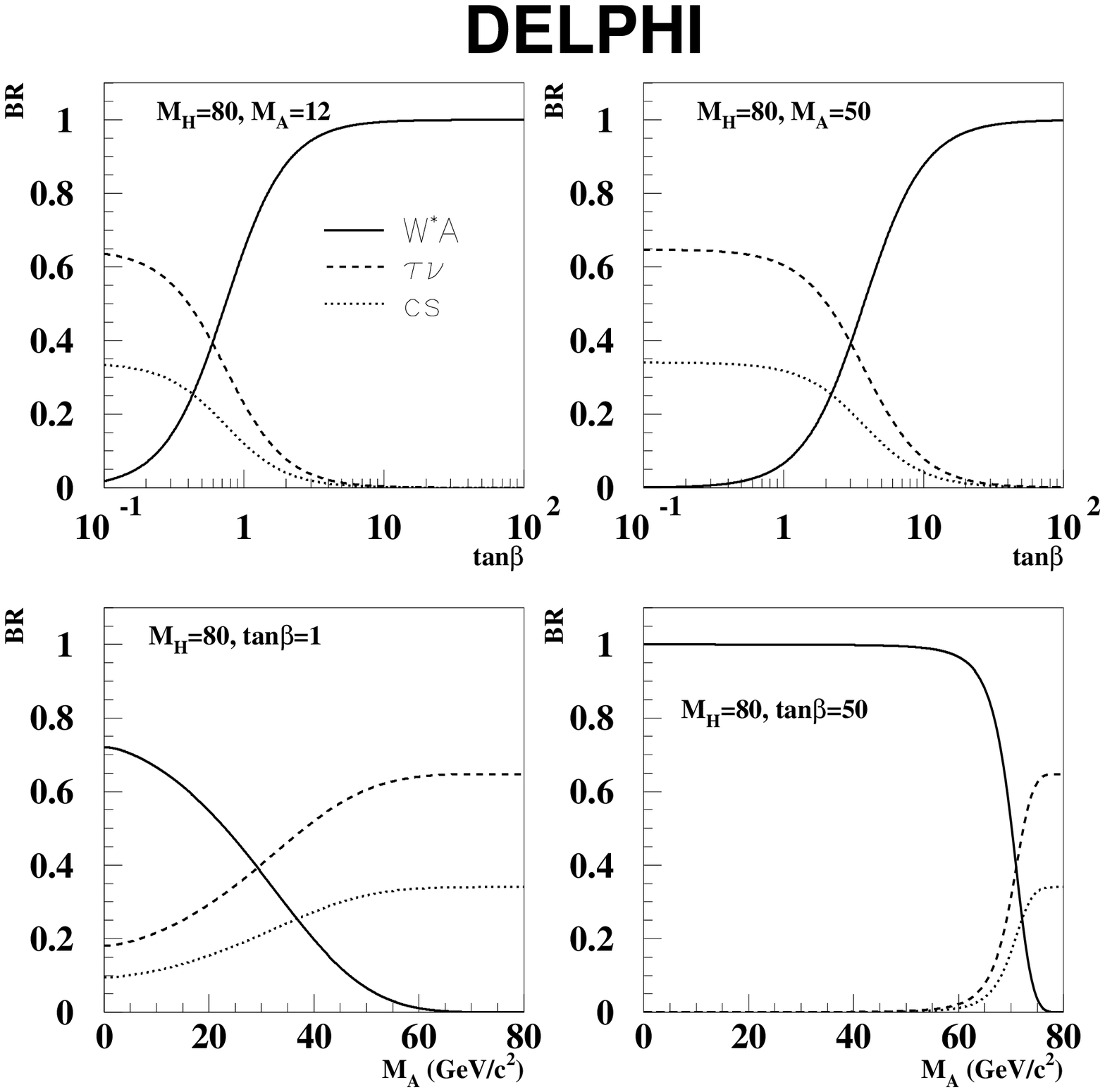,width=15.0cm}
\end{center}
\caption[]{Predicted charged Higgs boson decay 
branching ratios
for different parameters in the framework of type I Two Higgs Doublets Models.}

\label{fig:model}
\end{figure}

\begin{figure}[hbpt]
\begin{center}
\epsfig{file=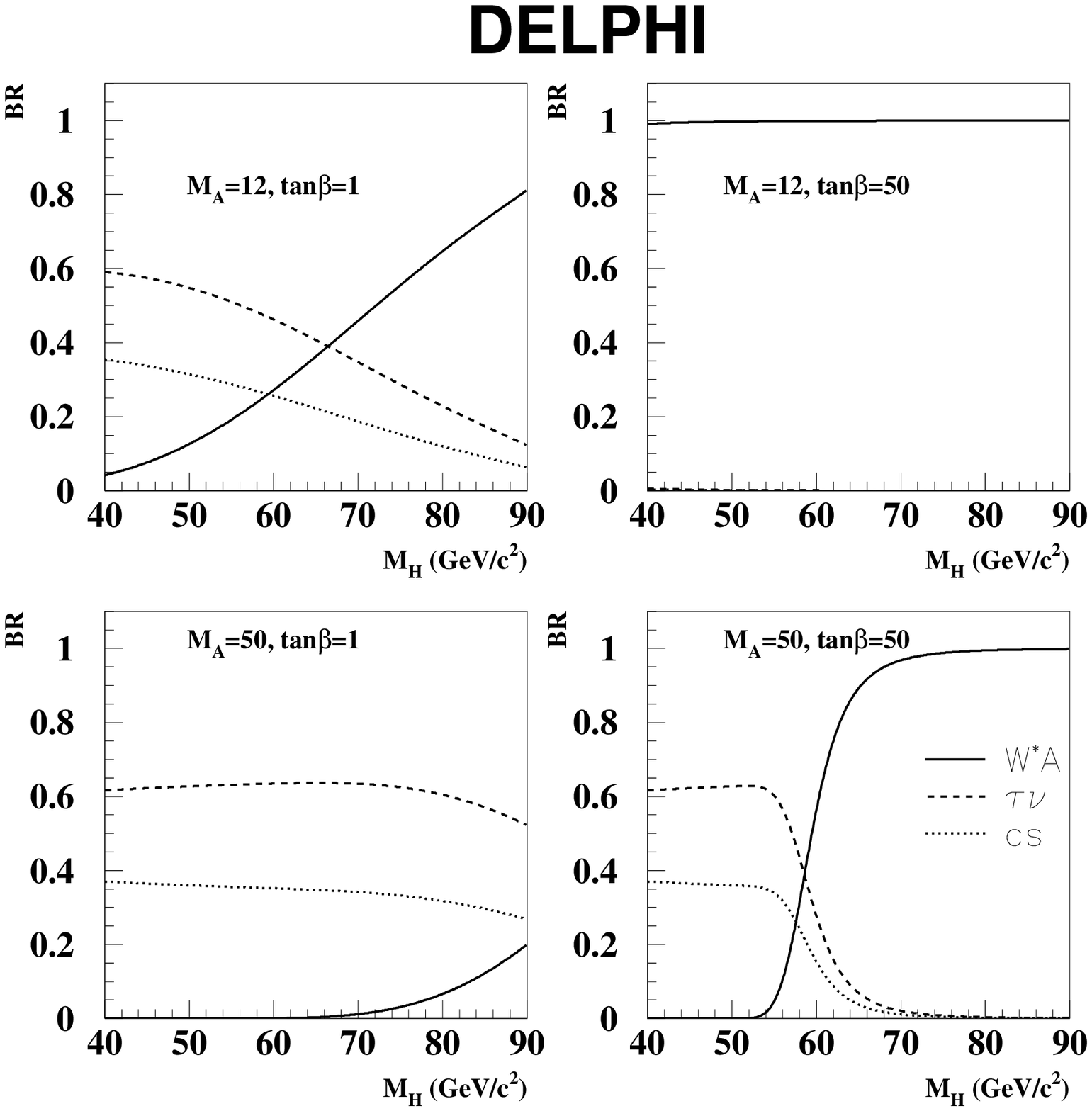,width=15.0cm}
\end{center}
\caption[]{Predicted charged Higgs boson decay 
branching ratios
for different parameters in the framework of type I Two Higgs Doublets Models.}

\label{fig:model2}
\end{figure}

The existence of a 
pair of charged Higgs bosons
is predicted by several
extensions of the Standard Model. Pair-production of charged Higgs
bosons occurs mainly via $s$-channel exchange of a photon or a $\zo$
boson. In Two Higgs Doublet 
Models (2HDM),
the couplings are completely specified
in terms of the electric charge and the weak mixing angle, $\theta_W$,
and therefore the production cross-section depends only on the charged
Higgs boson mass. Higgs bosons couple to mass and therefore decay
preferentially to heavy particles, but the precise 
branching ratios
may vary
significantly depending on the model. 
In most cases, for the masses accessible at LEP energies, 
the  $\tn$ and $\cs$ decay\footnote{Here and in the following all the decay modes are referred to the 
${\rm H}^-$, the charge conjugated being in all cases considered.} 
channels are expected to 
dominate. This is the case of the so-called type II 2HDM Models~\cite{typeII}, where one Higgs doublet 
couples
only to up-type fermions and the other to down-type fermions.
Analyses of the three possible final states, $\HTT$, $\HCSCS$ and $\HCSTN$, have been performed and are
described in this paper. To avoid loss of generality, the results are combined
and interpreted treating the Higgs decay 
branching fraction to leptons as a free parameter.
An alternative set of models, type I models~\cite{typeI}, assume that all fermions couple
to the same Higgs doublet. In this case and if the neutral
pseudo-scalar A is light 
(which is not
excluded by direct searches 
for
general 2HDM~\cite{delphi2hdm})
the decay to $\wa$ can be predominant even in the range of masses of interest
at LEP ($\mathrm{W}^*$ is an off-shell W boson). Figs.~\ref{fig:model} and~\ref{fig:model2} show the 
branching ratios
for different parameters 
in type I
models~\cite{WAteo}.
To cover the possibility of a light A boson 
the final states $\HWAWA$ and $\HWATN$ were also looked for. 
The channel $\HWACS$ is not considered because its contribution is
expected to be small. 
Type I models are explored through the combination of all the five channels, with or without $\wa$ decays. The
combination is performed according to the branching ratios predicted by the model as a
function of the ratio of the Higgs vacuum expectation ($\tan\beta$) and the boson masses.

Previous studies~\cite{delphihhlep} exclude masses below 43.5 \gev$/c^2$, for type II models. The limit is also
valid for type I models when the $\wa$
decay is not kinematically allowed or its branching ratio is small. 
Electroweak precision measurements~\cite{lepew} set indirect bounds on the charged Higgs
mass regardless of its decay branching ratios. The tree-level decay amplitude $\Gamma(\zo \to \hphm)$ is
independent of the model assumptions and can be calculated within 2HDM to be\cite{physlep}
\begin{equation*}
\Gamma(\zo \to \hphm)=\frac{G_F {\mz}^3}{6 \sqrt{2}\pi}{\left(\frac{1}{2}-{\sin^2\theta_W} \right)}^2 
{\left(1-{\frac{4 \mhp^2}{\mz^2}}\right)}^\frac{3}{2},
\end{equation*}
where ${\rm G_F}$ is the Fermi coupling constant, $\mhp$ and $\mz$ are 
the masses of the charged Higgs and $\zo$ and 
$\sin\theta_{\rm W}$ is the weak mixing angle.
The difference between the measured decay width of the $\zo$ ($\Gamma_{\mathrm{Z}}$) and the prediction from the 
Standard Model sets a limit to any non-standard contribution to the $\zo$ decay. The current 
results~\cite{lepew} set the limit 
$\Gamma_{non SM}<2.9$ MeV$/c^2$ at 95\% C.L. (taking into account both experimental and
theoretical errors),
which combined with the above expression sets
the limit $\mhp>39.6$ \gev$/c^2$ at  95\% C.L.
As a consequence, the searches in this analysis  are performed for charged Higgs boson masses of 
38 $ \gev/c^2$ or larger. 
 The limits described here are only valid if the neutral pseudoscalar is heavy enough to allow the $\bb$
decay.

Different techniques were developed to optimise the background rejection. In particular all analyses used
multidimensional estimators, likelihood functions or neural networks, to improve the discrimination. At the final
stage of the analyses the irreducible background from $\ww$ or $\zz$ events was discriminated by the inclusion of
specific variables such as the boson production angle, jet flavour tagging (both c/s and b-quark)
or $\tau$ polarisation.

\section{Data sample}

Data collected during the 2000 LEP run at centre-of-mass energies from 200 GeV 
to 209 GeV were used, with a total integrated luminosity of about 
220 pb$^{-1}$. The data were grouped into two samples with centre-of-mass energies 
      above or below 205.5 \gev, respectively. In the following, the average energy
      is quoted for each of these two samples.
Approximately 60  pb$^{-1}$ of these data were collected when one of the sectors of
the Time Projection Chamber (TPC) was not operational 
(referred to as the S6 period in the following). 
The data collected during the years 1998 and 1999 at centre-of-mass
energies from 189 GeV to 202 GeV were reanalysed, to take advantage of the
improved performance of the reconstruction and selection. 
The additional data amounted to approximately 380~pb$^{-1}$.  

The
DELPHI detector and its performance have already been described in
detail elsewhere~\cite{delphidet1,delphidet}.

The background estimates for the different Standard Model processes
were based on the following event generators: KK2f~\cite{kk2f} for
$\qq(\gamma)$ and $\mu^+\mu^-(\gamma)$, KORALZ~\cite{koralz} for
$\tau^+\tau^-(\gamma)$, BHWIDE~\cite{bhwide} for $e^+e^-(\gamma)$ and
WPHACT~\cite{wphact} for four-fermion final states. The four-fermion samples
were complemented with two-photon
interactions, generated with TWOGAM~\cite{twogam} for hadronic
final states, DIAG36~\cite{bdk} for electron final states and
\mbox{RADCOR}~\cite{bdk} for other leptonic final states.
The quark hadronisation was simulated with JETSET 7.4~\cite{pythia} and comparisons were made with HERWIG~\cite{herwig}
and ARIADNE~\cite{ariadne}.
All the relevant background were simulated at each of the main centre-of-mass energies 
with an equivalent luminosity of at least 40 times that recorded for the data.

Signal samples were simulated using the HZHA generator~\cite{HZHA} for charged Higgs masses from 40 to 100 \gev$/c^2$ in steps 
of 10~\gev$/c^2$, with additional points 
at 75, 85 and 95~\gev$/c^2$. For decays involving a neutral pseudoscalar, its mass
was varied from 20~\gev$/c^2$ up to the charged Higgs mass, with the same step size, with additional points at 
12~\gev$/c^2$. Samples of 2000 events were simulated for each mass point
for each of the five decay channels at the same centre-of-mass energies. The
$\mathrm{W}^*$ and $\mathrm{A}$ bosons, if present, were
allowed to decay according to the Standard Model and Two Higgs Doublet Model expectations, respectively.

A specific simulation, with the appropriate detector conditions, was performed for the S6 period, both for signal and
for background. 

\section{Analysis}

Most of the techniques and requirements follow closely those used
for the selection of $\ww$ pairs~\cite{delphiww}, since the topology of the $\hphm$
signal is very similar. We briefly describe them here together
with other techniques specific to the present analyses.

\subsection{Run selection and particle selection}
To ensure a good detector performance the data corresponding to runs in which
subdetectors relevant to the analysis were not fully operational were discarded. 
In particular it was
required
that the tracking subdetectors and calorimeters were fully
operational. 
An exception for the
S6 period was made, because the redundancy of the tracking system of the DELPHI detector made possible the use of 
these data
without a significant degradation of the analyses. 
For all the topologies that involved leptons, it was 
further required
that the muon chambers were active. This resulted in slightly 
smaller integrated luminosities than 
for
the hadronic channel. 
Table~\ref{tab:lumi} summarizes the luminosities selected in each case at every centre-of-mass energy.

\begin{table}[ht!]
\begin{center}
\begin{tabular}{ccc}
\hline
 $\sqrt{s} $(GeV) 
& \multicolumn{1}{c}{$\cal{L}$ (leptonic)}
& \multicolumn{1}{c}{$\cal{L}$ (hadronic)}\\
\hline
 189       & 153.8  &  158.0\\
 192       &  24.5  &  25.9\\
 196       &  72.4  &  76.9\\
 200       &  81.8  &  84.3\\
 202       &  39.4  &  41.1 \\
 205       &  69.1  &  75.6\\
 206.6     &  79.8  &  87.8\\
 206.3(S6) &  50.0  &  60.8\\
\hline
\end{tabular}
\end{center}
\caption[]{Integrated luminosity in pb$^{-1}$ selected for leptonic and hadronic final states at the
different centre-of-mass energies.} 

\label{tab:lumi}

\end{table}


Only charged particle tracks with an impact parameter in the transverse plane\footnote{The co-ordinate
  system used has the $z$-axis parallel to the electron beam, and the
  polar angle calculated with respect to this axis. The plane perpendicular to the $z$ axis 
  will be called hereafter the transverse plane.} smaller than 5 cm, and
with an axial coordinate $|z| < 10~\mbox{cm}$ at the point of closest approach to the
beam spot, were accepted.
Those with a 
relative momentum error $\frac{\Delta p}{p}>1$ 
were rejected.

Showers in the calorimeters were accepted as neutral particles 
if their energy was above 200 MeV\@ and they were not associated to a charged particle track.

\subsection{Lepton identification}

To perform lepton identification, an initial clustering of particles into jets was performed
with the LUCLUS~\cite{pythia} algorithm.

Jets containing only one charged particle and no neutral particles, which were isolated by
more than $15^\circ$ from the remaining particles in the event, were initially
considered as lepton candidates.
One of these isolated charged particles was  identified as a muon if it gave signal in the muon
chambers or left a signal in the calorimeters compatible with a minimum ionising particle (MIP\@).
It was identified as an electron if its energy deposition in the
electromagnetic calorimeters was compatible with its measured momentum and the
ionisation loss in the TPC was compatible with that expected from an electron
of that momentum.

If an electron or muon had a momentum and energy deposition in the
electromagnetic calorimeters smaller than $0.13\sqrt{s}$, 
it was assumed to come from a
$\tau$ decay and 
was therefore
tagged as $\tau$. 
In addition,
isolated jets with an energy of at least 5 GeV, at least 
one and at most five charged particles and no more than ten
particles in total were also considered as $\tau$ candidates. 

When dealing with semileptonic final states, the $\tau$ candidate jet 
definition was refined removing particles that were not likely
to come from a $\tau$ decay. 
Particles contained inside the jet, but 
forming an angle with the
jet axis of more than
$15^\circ$ were 
removed
from the jet.
If the
invariant mass of the jet was greater than 
2.5 GeV$/c^2$, 
the particle giving
the largest contribution to the mass (excluding the highest momentum charged
particle in the jet) was excluded. This procedure was repeated
until the mass no longer exceeded 2.5 GeV$/c^2$.

If more than one $\tau$ candidate was found 
they were sorted with the following order of precedence:
muon, electron, narrowest jet (defined as the one whose momentum-weighted angular spread was lowest), single 
charged particle. When more than one leptonic or single charged particle candidates of the same type were found, they
were sorted according to the product of the measured energy and isolation angle\footnote{The isolation angle for a
particle defined as the minimum angle between that particle and any other charged particle in the event.}. 
For purely leptonic events the first two candidates were retained and the rest were neglected as $\tau$
particles. For semileptonic events, only the first one was retained as a $\tau$ candidate.

In some of the analyses, the particular decay of the $\tau$ had to be identified.
All $\tau$ candidates were classified into the following categories (corresponding to the major
decay modes):
$e$, $\mu$, $\pi$, $\pi+n\gamma$, $\geq3\pi$ according to the lepton
identification, the number of charged particles of the jet and the number of
photons.

\subsection{Likelihood ratio technique}

In several of the analyses the background discrimination was
performed  using a likelihood ratio technique. Signal and background 
likelihood functions, ${\cal L}_s$ and ${\cal L}_b$, were defined as products
of the probability density functions of the $N$ discriminating variables,
${\cal L}_s=\prod_{i=1,N} s_{i}(x_i)$ and ${\cal L}_b=\prod_{i=1,N}
b_{i}(x_i)$. For each of the measured values of the $N$ discriminating
variables, $x_i$, the values of the signal and background
probability densities, $s_{i}(x_i)$ and $b_{i}(x_i)$, were determined
using samples of simulated signal and background events. The final
event likelihood ratio, 
for simplicity referred to as ``likelihood'' in the following,
was computed as a normalised ratio of the
signal and background likelihoods, ${\cal L}_s/({\cal L}_s+{\cal
  L}_b$).
  
\subsection{Tau polarisation}

One of the methods used to discriminate charged Higgs from W bosons is 
based on the different spin of these particles,  the Higgs being a scalar and the W a vector boson. 
This spin can be inferred if the decay
involves $\tau$ leptons.

Assuming that the $\nu_\tau$ has a definite helicity, the polarisation
($P_\tau$) of tau leptons originating from heavy boson decays is
determined entirely by the properties of weak interactions and the
spin of the parent boson. The helicity configuration for the signal
is H$^- \rightarrow \tau^-_R \mbox{$\bar\nu_\tau$}^{\phantom +}_R$
(H$^+ \rightarrow \tau^+_L {\nu_\tau}^{\phantom +}_L$) and for the
$\ww$ background it is W$^- \rightarrow \tau^-_L
\mbox{$\bar\nu_\tau$}^{\phantom +}_R$ (W$^+ \rightarrow \tau^+_R
{\nu_\tau}^{\phantom +}_L$) resulting in $P_\tau^{\rm H}=+1$ and
$P_\tau^{\rm W}=-1$. 

The $\tau$ weak decay induces a 
dependence
of the angular and momentum distributions on the
polarisation. Once the $\tau$ decay channel was identified, the
information on the $\tau$ polarisation was extracted from the observed
kinematic distributions of its decay products, e.g. angles and
momenta.  These kinematic variables can always be combined~\cite{rouge} into a single estimator, defined for each decay
channel, without loss of information.
These estimators are equivalent to those used at the $\zo$ peak for
precision measurements~\cite{ptauz}. For charged Higgs boson masses close to the
threshold, the boost of the bosons is relatively small and the $\tau$
energies are similar to that of the $\tau$'s from 
$\zo$
decays at rest (40--50~GeV). 

To coherently compare events in which the $\tau$ had different decay modes, 
the identified decay mode and $P_\tau$ estimator were
combined into a likelihood function. When two $\tau$ candidates 
were present in one event, the likelihood functions
were defined for each of them and
then multiplied, assuming independence of the two $\tau$ (which was true to a 
large extent, except for some small correlations due to detector
effects). 

\subsection{Jet definition and flavour tagging}

When the charged particle multiplicity was larger than 6, 
the particles  were clustered into
jets using the DURHAM\cite{durham} algorithm.
When a $\tau$ had been identified, the particles assigned to its jet  were excluded from
this clustering and the remaining particles were forced into exactly two jets. 
Each of the two jets was required to
have a minimum of  four particles of which at least one had to be charged.
For the purely hadronic events, the jet algorithm was forced to produce
a maximum of four jets.


In the $\HCSCS$ and $\HCSTN$ decay channels all  hadronic jets
in the event originate from a c or s quark. In the hadronic
background processes, such as $\qq$ and $\ww$ events, often the
jets have a different quark flavour or originate from a gluon. Therefore
a jet flavour-tagging algorithm was used as a tool in the
analyses of the $\HCSCS$ and $\HCSTN$ channels. The algorithm follows a similar 
technique to that used by DELPHI in a determination of
$|V_{cs}|$ at LEP~II~\cite{jetag}. 

This tagging was based on nine discriminating variables, combined in a likelihood function: 
three of them
were related to identified leptons and the hadron content of the jet,
two depended on kinematical variables and four on the reconstructed
secondary decay structure. The finite lifetime of c-quark containing
particles was exploited to distinguish between c and light quark
jets, while the c-quark mass and decay multiplicity were used to
discriminate against b jets. Furthermore s and c jets could be
distinguished from u and d jets by the presence of an identified
energetic kaon. Charged hadrons were identified combining
the Ring Imaging Cherenkov (RICH) and TPC $dE/dx$~\cite{part_id} measurements. 
The responses of the flavour-tagging algorithm for the individual jets
were further combined into an event $\HCSCS$ probability or
into a di-jet $\cs$ (or $\csb$) probability which were then used in background
suppression.

\label{sec:cstag}

\subsection{Mass reconstruction}

The masses of the decaying bosons were reconstructed using a constrained
fit~\cite{pufitc} requiring energy and momentum conservation with known beam energy 
(4C fit).
For the topologies studied in this analysis, the event had to be compatible with the 
hypothesis that the different objects were
produced in the decay of two equal mass particles, so an additional constraint was
applied requiring that the two mass combinations were equal (5C fit).
These fits also provide the best estimation of the boson momenta.

In the case of channels involving a $\tn$ decay, the three 
components of the
momentum vector of the $\nu_\tau$ and the magnitude of the $\tau$
momentum were treated as unknown parameters, reducing the number of
degrees of freedom of the fit from five to one. 
This fit also provided a good estimation of the $\tau$ 4-momentum.

In the $\HTT$ final state, the number of unknowns was higher than the
number of constraints and no mass could be estimated.

\section{Selection}

\subsection{The $\HTT$ channel}

The signature for $\hphm \rightarrow \HTT$ is large missing energy and momentum and two
acollinear and acoplanar\footnote{The acoplanarity is defined as the
  supplement of the angle between the two jets projected onto the
  plane perpendicular to the beam.} $\tau$ jets containing either a lepton or
one or a few hadrons. The main backgrounds are the $\ww$ leptonic decays, mainly those
in which one W or both decayed to $\tau \nu$. Less important, but still not negligible,
are the radiative $\tto$ and two-photon events.

\subsubsection{Event preselection}

To select leptonic events a total charged particle multiplicity
between 2 and 6 was required. 
Only events with two jets
both containing at least one charged particle were retained. 
Events were rejected if both jets had
more than one charged particle.  
It was also required that the
angle between the two jets was larger than $30^\circ$.

Two-fermion and two-photon events were rejected 
by requiring the acoplanarity to be
larger than $13^\circ$ if both jets were in the
barrel region ($43^\circ<\theta<137^\circ$) or larger than $25^\circ$ otherwise.


The two-photon background was further reduced by different requirements 
on the jets:
the sum of the jet energies transverse to the
beam direction, $E_\perp$, was required to be greater than 
$0.1 \sqrt{s}$; 
the total transverse momentum,
$P_T$, to be greater than $0.04 \sqrt{s}$; the 
total energy detected within $30^\circ$ around the beam axis to be less
than $0.1 \sqrt{s}$; and the total energy outside this region to be greater than 
$0.1 \sqrt{s}$.

To reject $\ww$ events where neither W 
decayed to $\tau \nu$, 
it was required
that the two jets were identified as $\tau$ leptons.


\subsubsection{Final background discrimination}

Following the selection above
most of the remaining background is \mbox{$\ww \rightarrow \HTT$} events. 
The $\hphm$ signal and the $\ww$ background have
similar topologies and the presence of missing neutrinos in the
decay of each of the bosons makes  the boson mass reconstruction  impossible.
However, the boson polar angle distribution and the $\tau$ polarisation are different, 
providing means to discriminate between the two processes.

\begin{figure}[hbpt]
\begin{center}
\epsfig{file=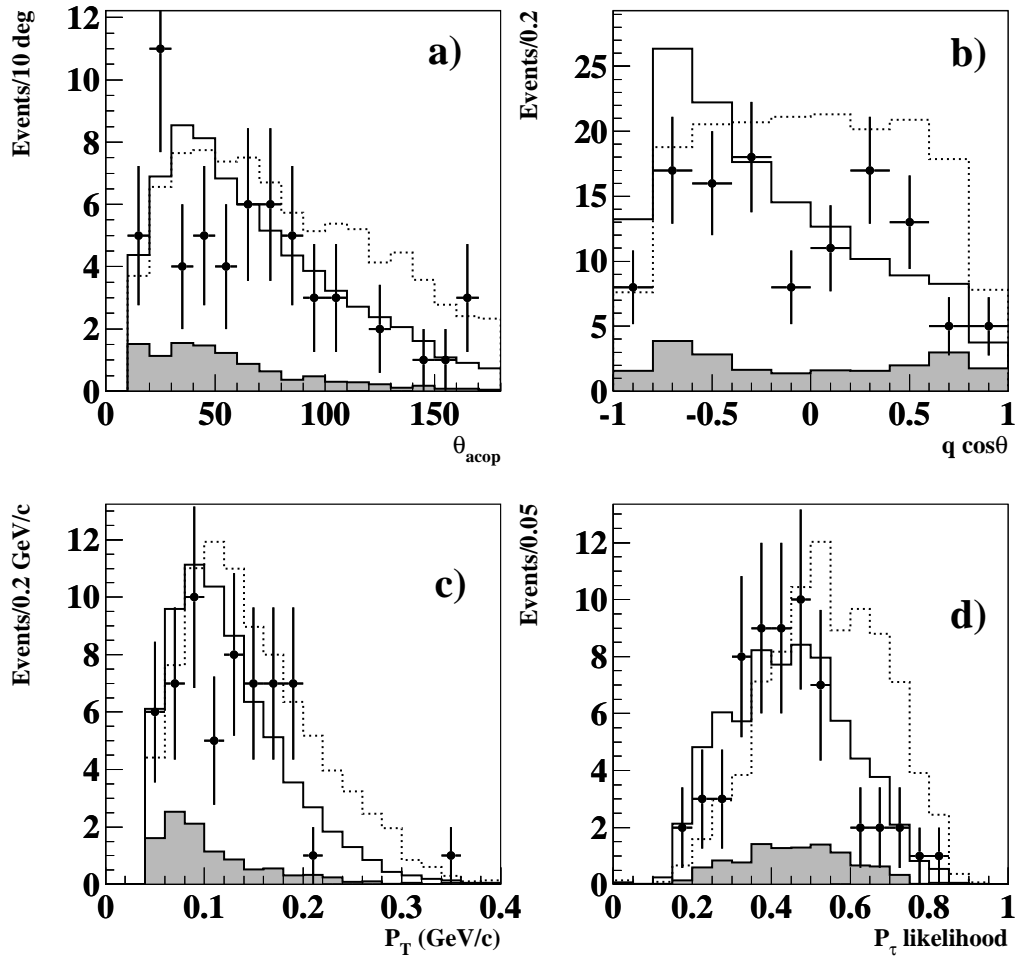,width=15.0cm}
\end{center}
\caption[]{Distribution of some of the variables used for the anti-WW likelihood
     for the $\HTT$ analysis
  at $\sqs=$ 189--209~GeV\@ after preselection: 
  a) acoplanarity, b) signed cosine of polar angle of each $\tau$ candidate,
  c) 
 total transverse
momentum and d) event $\tau$ polarisation likelihood.
  Data are shown as filled circles, while
  the solid histogram contour shows the expected SM 
  background with contributions from $\ww$  (unfilled) and $\qq$ (shaded).
  The expected histogram for a
  85 GeV/$c^2$ charged Higgs boson signal is shown as a dashed line
  with arbitrary normalisation
  for comparison.}

\label{fig:vartautau}
\end{figure}

\begin{figure}[hbpt]
\begin{center}
\epsfig{file=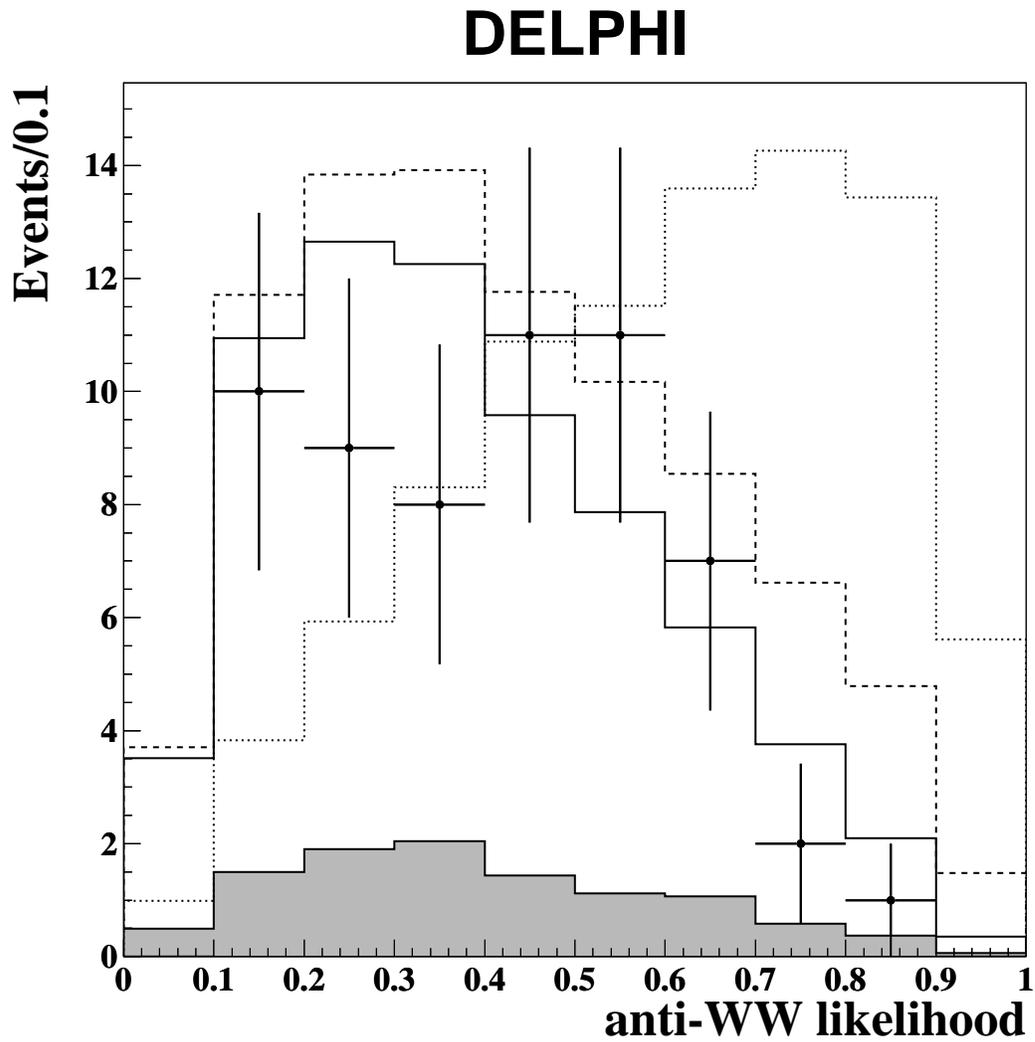,width=15.0cm}
\end{center}
\caption[]{Distribution of the anti-WW likelihood
  for the $\HTT$ analysis 
 at $\sqs=$189--209~GeV\@. 
 The dots represent the data, 
 while the solid histogram contour shows
 the expectation from SM 
  processes, as in Fig. \ref{fig:vartautau}.
  The expected histogram for a
  85 GeV/$c^2$ charged Higgs boson signal has been normalised to the
  production cross-section and 100\% leptonic branching ratio and
  added to the 
  backgrounds (dashed). 
  The dotted line shows the shape of the likelihood for the
  charged Higgs signal only in arbitrary normalization.}

\label{fig:lik80}
\end{figure}

A likelihood function was built to separate the signal from the background.
It was composed of five variables: the $\tau$
polarisation likelihood of the event, the signed cosine of the polar 
angle\footnote{The signed cosine is defined as the
charge of the particle multiplied by the cosine of its polar angle.} of 
both $\tau$'s (which carries information of the boson polar
angle), the acoplanarity and the 
total transverse momentum. The first three variables discriminated between 
$\HTT$  produced from W boson and charged Higgs pairs. The last two variables
had some sensitivity to the boson mass and 
helped in the discrimination of the remaining background from other processes.
Some of these variables are
shown in Fig.~\ref{fig:vartautau} and the resulting 
likelihood distribution for data, expected backgrounds and
signal is shown in Fig.~\ref{fig:lik80}. The effects of the different
sets of cuts are shown in Table~\ref{tab:tntnsel} for the combined
$\sqs=$189--209~GeV sample. 

\begin{table} [ht!]
\begin{center}
\vspace{1ex}
\begin{tabular}{lrrrrr}
\hline
cut & \multicolumn{1}{c}{data}
    & \multicolumn{1}{c}{total bkg.}
    & \multicolumn{1}{c}{4-fermion}
    & \multicolumn{1}{c}{other bkg.}
    & \multicolumn{1}{c}{$\varepsilon_{80}$} \\
\hline

Leptonic selection    & 175699 & 176685  &   921 &  175764  &    72.2\%
\\
Acoplanarity cut      &  16607 &  16576  &   715 &  15861   &   62.3\%
\\
Energy/momentum cuts   &    527 &    566.9  &   534.4 &     32.5   &   46.7\%
\\
$\tau$ identification &     59 &     68.9  &    58.3 &     10.6   &   35.1\%
\\
\hline

\end{tabular}
\end{center}
\caption{The total number of events observed and expected backgrounds
  in the $\HTT$ channel after the different cuts used in the
  analysis at $\sqs=$ 189--209~GeV\@. The last column shows the efficiency for a charged Higgs
  boson signal with $\mhp = 80$~GeV/$c^2$.}
\label{tab:tntnsel}
\end{table}

\subsection{The $ \HCSCS$ channel}

In the analysis of 
the $\HCSCS$
channel both charged Higgs bosons were 
assumed
to decay into a pair of $\mathrm{c}$ and $\mathrm{s}$ quarks producing a final state
with four jets. The two 
dominant background
sources  are the
$\qq$ production with gluon radiation ($\qq gg$) and fully hadronic four-fermion final
states. 
The four-fermion background from $\ww$ production is much more severe than that
from $\zz$, because of the higher cross-section.
In addition, the discriminant variables used against the $\ww$  background usually work with
similar performance against the $\zz$ background.  Therefore, the
four-fermion sample is referred to as $\ww$  in 
the remainder of the section.

\subsubsection{Event preselection \label{subsub:evtpresel}}


In order to preselect hadronic events the following cuts were applied:
the events had to contain at least 10 charged particles, the visible energy of the 
reconstructed
particles, 
 had to be larger than $0.6 \sqrt{s}$, the reconstructed
effective centre-of-mass energy\footnote{The effective centre-of-mass energy was estimated from a three-constraint
kinematic fit in order to test the presence of an initial state radiated photon lost in the beam
 pipe~\cite{sprime}.}, $\sqrt{s'}$, 
had to be larger than $0.85
\sqrt{s}$. To reject hadronic back-to-back two-jet $\qq$ events
the thrust was required to be less than 0.95.
  
To select only
genuine four-jet events it was required that the DURHAM clustering 
distance 
for the transition from four to three jets, $y_{4 \rightarrow 3}$, 
was
greater than 0.002 and  each jet was required to have a mass larger 
than 2 GeV/$c^2$ and at
least three 
particles, out of which at least two were charged. All jets were required
to have a total energy above 5 GeV, and the minimum angle between 
any two
jets was required to be at least 25$^\circ$.

In order to obtain the best possible mass resolution 
a 5C fit was performed for each of the three possible 
di-jet combinations and the combination giving the smallest
$\chi^2$ was selected. 
A 4C
fit was also performed for that di-jet combination, imposing only energy-momentum conservation, to estimate the difference between the masses 
of the two reconstructed bosons. 
As the uncertainty of
the di-jet mass reconstruction is
approximately proportional to the mass,
the boson mass difference was renormalised 
dividing it by the mass provided by the 5C fit.
In this a way, the resulting discriminant variable had
less dependence on the signal mass. 
This relative
mass difference of the two reconstructed bosons was required to
be below 25\%.


\subsubsection{Final background discrimination}

Significant amounts of $\qq g g$ background still remained after
the preselection. To suppress it further an anti-$\qq$ likelihood
function based on five variables was constructed as follows.

\begin{figure}[hbpt]
\begin{center}
  \epsfig{file=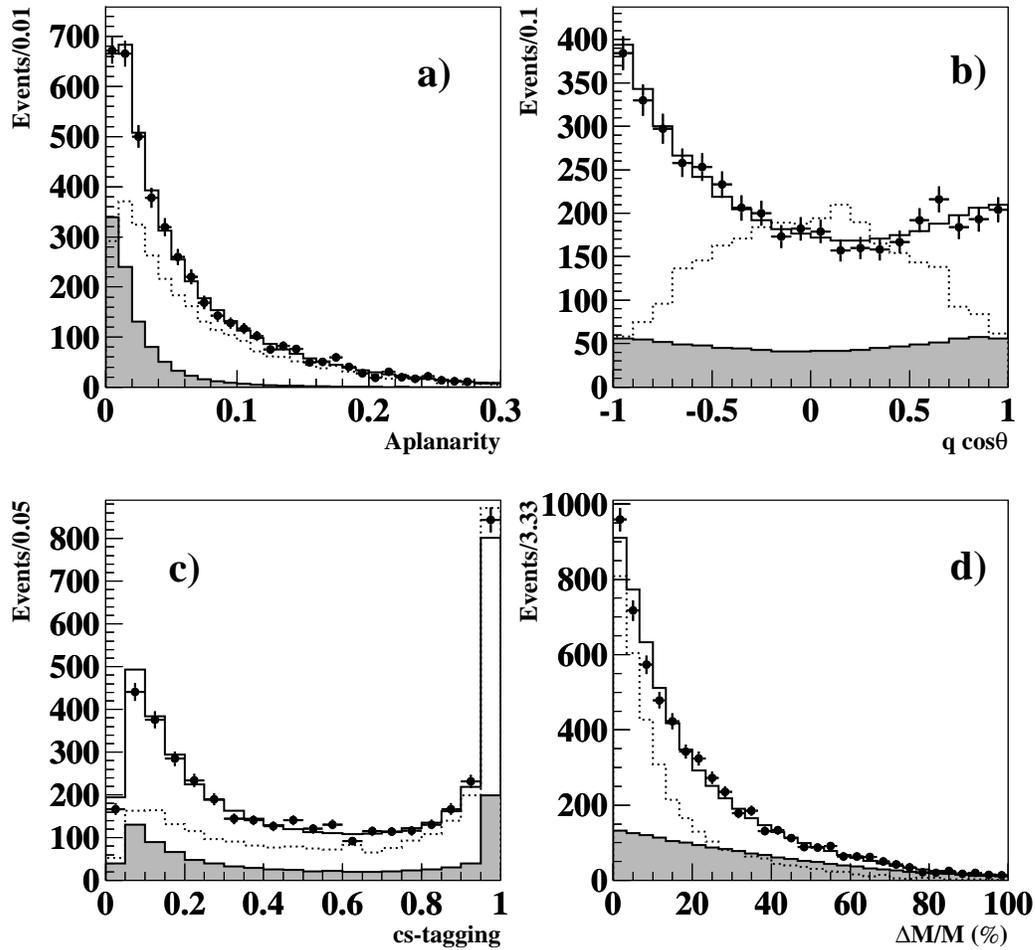,width=15.0cm}
\end{center}
\caption[]{Distribution of some of the variables used for the 
  anti-$\qq$  and anti-WW likelihoods
  in the $\HCSCS$ analysis
  at $\sqs=$189--209~GeV\@ after preselection: 
  a) aplanarity, b) signed cosine of the polar angle of the boson
  c) $\cs$-tagging variable and d) relative mass difference.
  Data are shown as filled circles, while
  the solid histogram shows the expected SM 
  background with contributions from $\ww$  (unfilled) and $\qq$ (shaded).
  The expected 
 distribution
  for a
  75 GeV/$c^2$ charged Higgs boson signal is shown as a 
  dotted histogram with
  arbitrary normalisation
  for comparison.}

\label{fig:varscscs}
\end{figure}

The first variable, the event
aplanarity~\cite{jetsetman}, exploits the differences in
the event shape between signal and background. 
The second one, the cosine of the polar angle of the
thrust axis, uses the fact that the signal events have a polar angle
distribution with an approximate dependence as ${\sin^2\theta}$, 
whereas the jets in the $\qq$ background events are
concentrated closer to the beam axis.  
The third variable was based on the product of the
minimum angle between two jets and the minimum jet energy in the event and
exploited the particular dependence 
of the probability of hard gluon radiation with the
gluon energy and emission angle.
The minimum energy and the minimum angle between jets are
significantly different in signal events with low and high mass due to
the large boost of light Higgs bosons. In order to reduce the mass
dependence of the likelihood variable, the product 
was scaled dividing it by the reconstructed
Higgs boson mass of the event. 
The fourth variable used the 
fact that
the charged Higgs bosons have equal mass whereas the masses of the
di-jet pairs in the $\qq$ events are more or less randomly
distributed. Therefore the relative mass difference
was a powerful discriminant variable.
The last variable  was the output of the
event $\HCSCS$-tag described in Section~\ref{sec:cstag}.
The normalised likelihood was required to exceed 0.4
to reject most of the $\qq$ background with a moderate signal efficiency loss (Table~\ref{tab:cscssel}).
Some of these variables are shown in Fig.~\ref{fig:varscscs}.

Most of the background remaining after the anti-$\qq$ cut was hadronic
decays of W pairs. If the mass of the charged Higgs boson
coincides with the mass of the W boson the $\ww$ background is partly
irreducible. Some differences, however, exist and were combined into an anti-WW likelihood in order
to discriminate between these two processes. 

The first of the variables in the anti-WW likelihood exploited the
different polar angle distributions of the Higgs boson and the W boson, 
due to their different spins. This variable was the cosine of the polar angle of the 
       positive boson, estimated assuming equal and opposite boson momenta.
The charge 
was derived from
the sum of the momentum-weighted charges of the two
jets~\cite{TGC} 
used to reconstruct the boson.
The boson with the higher value of
charge was assumed to be the positive one and the other was
assumed to be the negative one. The second variable used for $\ww$
background discrimination was the $\HCSCS$ event tag output which is
useful as all signal jets originate from c and s quarks and only
half of the background jets have the same quark flavours. The last
variable used was the relative mass difference between the two
reconstructed bosons.  This variable has rejection power especially in
cases where the reconstructed mass 
in
W events is far away from
the nominal W mass since in these events something has gone wrong in the
jet momentum measurement, which usually leads to a higher mass
difference between the reconstructed bosons. It also rejects more $\ww$
background than charged Higgs signal due to a larger natural width of
the W boson.  
All
events with anti-WW likelihood value below
0.3 were rejected.

\begin{figure}[hbpt]
\begin{center}
  \epsfig{file=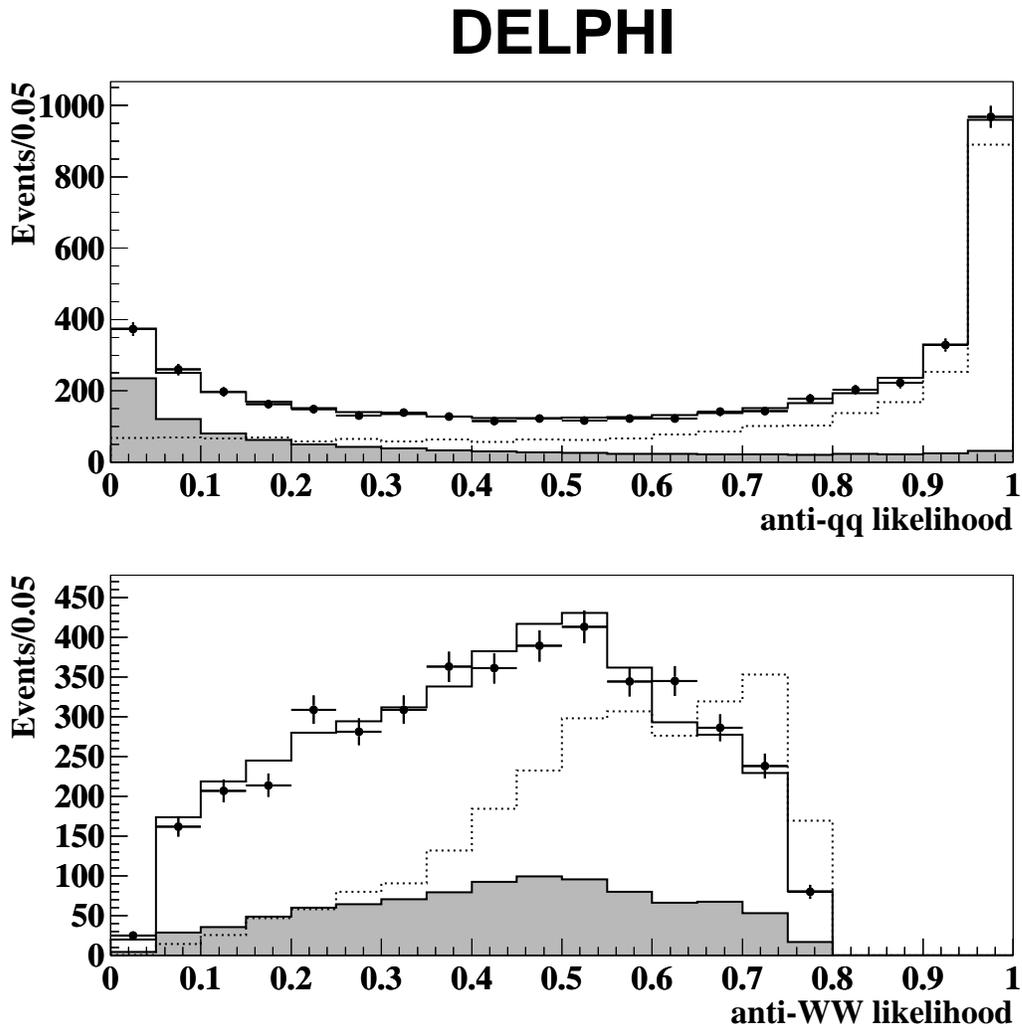,width=15.0cm}
\end{center}
\caption[]{Distributions of the anti-$\qq$  (top) and anti-WW (bottom) likelihoods
     in the  $\HCSCS$ analysis
  at $\sqs=$189--209~GeV\@ after preselection and mass difference cut. 
  Data and expected SM backgrounds are indicated as in Fig. \ref{fig:varscscs}.
 The expected distribution
  for a 75 GeV/$c^2$ charged Higgs boson signal is shown as a 
  dotted histogram with
  arbitrary normalisation.}
\label{fig:cscslike}
\end{figure}

\begin{figure}[hbpt]
\begin{center}
  \epsfig{file=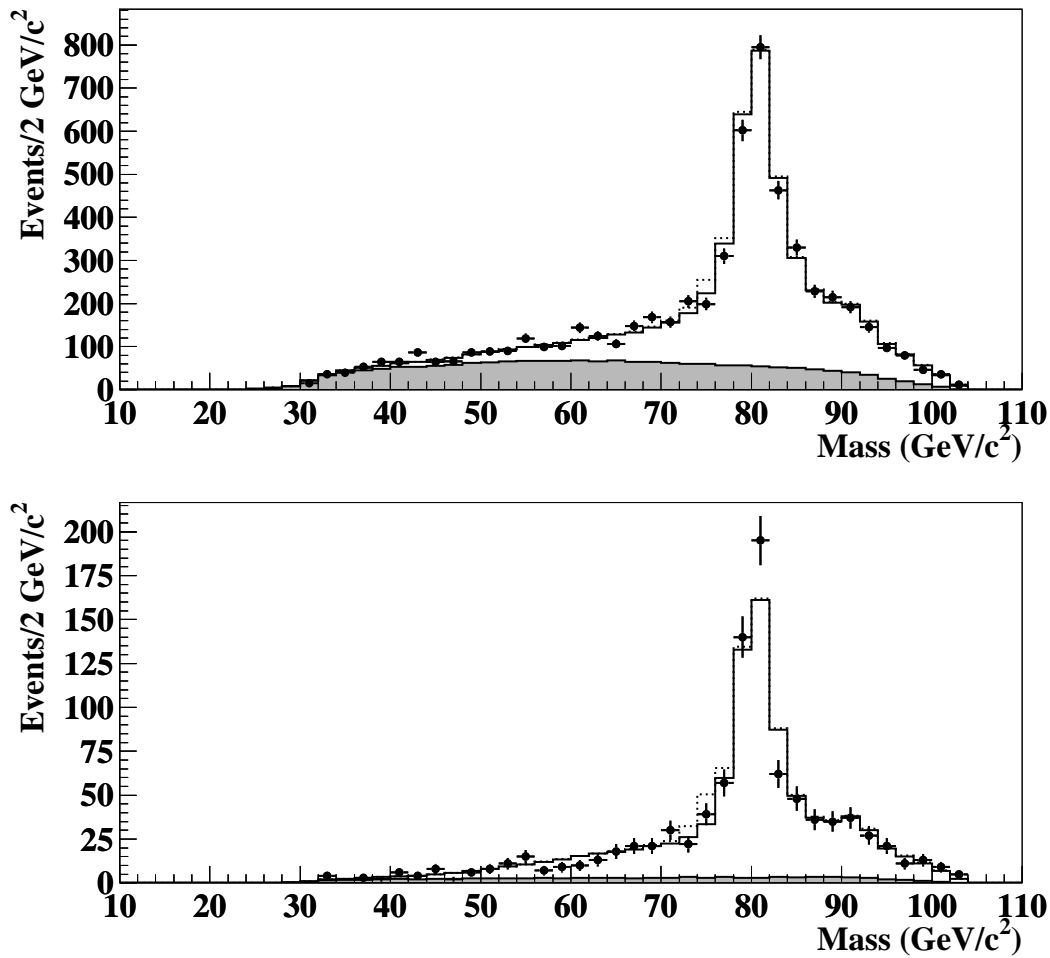,width=15.0cm}
\end{center}
\caption[]{Reconstructed mass distribution 
  in the  $\HCSCS$ analysis 
  at 
  $\sqs=$ 189--209~GeV at preselection (top) and
  at the final selection with additional cut of
   ${\cal L}_{\mathrm{qq}}>0.7$ and
   ${\cal L}_{\mathrm{WW}} > 0.5$ (bottom).
  Data are shown as filled circles, while
  the solid histogram shows the expected SM 
  background with contributions from $\ww$  (unfilled) and $\qq$ (shaded).
   The expected distribution in the presence of an 
   H$^+$H$^-$ signal, with $\mhp = 75~\mbox{\rm GeV}/c^2$
   and hadronic branching ratio of 100\%, is also shown for 
   comparison (dotted).}
\label{fig:cscsmass}
\end{figure}

The effects of the different sets of cuts are shown in
Table~\ref{tab:cscssel} for the combined $\sqs=$189--209~GeV sample. The
distribution of the anti-$\qq$ and anti-WW likelihoods
at
the preselection level 
are shown in Fig.~\ref{fig:cscslike}. The reconstructed 5C fit mass
distribution for data, expected backgrounds and signal after the 
anti-$\qq$ and anti-WW cuts is shown in Fig.~\ref{fig:cscsmass} with the likelihood cut tightened to
${\cal L}_{\mathrm{qq}}>0.7$ and 
${\cal L}_{\mathrm{WW}} > 0.5$ to visually enhance the mass distribution of the events whose variables are closer
to those expected for the charged Higgs signal.

\begin{table} [ht!]
\begin{center}
\vspace{1ex}
\begin{tabular}{lrrrrrr}

\hline
cut & \multicolumn{1}{c}{data}
    & \multicolumn{1}{c}{total bkg.}
    & \multicolumn{1}{c}{4-fermion}
    & \multicolumn{1}{c}{other bkg.}
    & \multicolumn{1}{c}{$\varepsilon_{75}$}  
    & \multicolumn{1}{c}{$\varepsilon_{80}$} \\ 
\hline
4-jet presel.      & 5890 &   5902.5 &   4076.9 &   1825.6& 83.0\% & 84.1\% \\
Mass diff.         & 4326 &   4354.2 &   3389.6 &    964.6& 71.0\% & 71.8\% \\
anti-$\qq$            & 2785 &   2808.1 &   2506.2 &    301.9& 56.9\%  & 57.8\% \\
anti-WW            & 2114 &   2115.6 &   1855.5 &    260.1& 52.8\%  & 53.6\% \\
\hline

\end{tabular}
\end{center}
\caption{The total number of events observed and expected backgrounds
  in the $\HCSCS$ channel after the different cuts used in the
  analysis at $\sqs=$ 189--209~GeV\@. The last 
  columns
  show the 
  efficiencies for charged Higgs boson signals
  with $\mhp = 75$~GeV/$c^2$ and $\mhp = 80$~GeV/$c^2$, respectively. 
}
\label{tab:cscssel}
\end{table}

\subsection{The $ \HCSTN$ channel}

In the  $\HCSTN$
channel one of the charged Higgs bosons decays into a $\csb$ 
quark pair, while the other decays into $\tn$. Such an
event is characterised by two hadronic jets, a $\tau$ candidate and
missing energy carried by the neutrinos. The dominant background
processes are $\qq g(\gamma)$ events and semileptonic decays
of $\ww$. 

\subsubsection{Event preselection}

An initial set of cuts was applied to reject purely leptonic events as well as events from two-photon
interactions. 
The charged particle multiplicity had to be at least 6
and the total momentum of the
charged particles had
to be greater than $0.01\sqrt{s}$. 
The quantity $ {\rm E_{fw}=\sqrt{{E_{45}}^2+{E_{135}}^2}}$, where ${\rm E_{45}}$ 
and ${\rm E_{135}}$ are the energies deposited in the electromagnetic calorimeters
at $\theta < 45^\circ$ and $\theta > 135^\circ$ respectively, had to be less than
$0.45\sqrt{s}$. The absolute value of the
cosine of the polar angle of the missing momentum
had to be less than 0.985 and the total transverse energy
had to be greater
than $0.2\sqrt{s}$. The electromagnetic energy within a $15^\circ$
cone around the beam-pipe was required to be less than 30 GeV\@.

To remove $\qq \ell^+\ell^-$  four-fermion topologies, events with two or more 
leptons of the same flavour with
momentum greater than $0.05\sqrt{s}$ and more than $10^\circ$ isolation
angle were rejected.


Another set of cuts was applied to reject the bulk of the 
$\qq\gamma$ radiative events.
The absolute value of the
cosine of the polar angle of the missing momentum
had to be less than 0.96, the
difference between the centre-of-mass energy and the effective 
centre-of-mass energy ($\sqrt{s} - \sqrt{s'}$) had to be greater than 10 GeV, and
the visible energy had to be lower than $0.85\sqrt{s}$. The
DURHAM clustering distance 
$y_{4\rightarrow 3}$
had to be less than 0.03.
The angle between the most energetic neutral particle in the event and
the missing momentum had to be greater than $25^\circ$.
If the absolute value of the cosine of the polar angle of 
the missing momentum was greater than 0.8, the
effective centre-of-mass energy ($\sqrt{s'}$) had to be greater than 105
GeV and its difference from the nominal centre-of-mass energy ($\sqrt{s} -
\sqrt{s'}$) had to be greater than 25 GeV\@.

Background from $\ww$ semileptonic decays not involving $\tau$ particles as well as a large fraction
of the remaining $\qq$ background was rejected by requiring the presence of an identified $\tau$.
The momentum of the $\tau$ jet had to be greater than 5 \gev$/c$
and the product of the $\tau$ candidate momentum and its isolation angle
had to be larger than 150 GeV$\cdot$degree. If the $\tau$
candidate jet contained more than one charged particle, the cone around its
axis containing 75\% of the jet energy had to be smaller than
$10^\circ$.


Finally, if the 5C mass fit  did not
converge the event was rejected. This reduced the background from misreconstructed $\ww$  pairs, 
with badly defined jets or
with wrong pairing, contributing to masses very different from the expected W peak.

\subsubsection{Final background discrimination}

\begin{figure}[hbpt]
\begin{center}
  \epsfig{file= 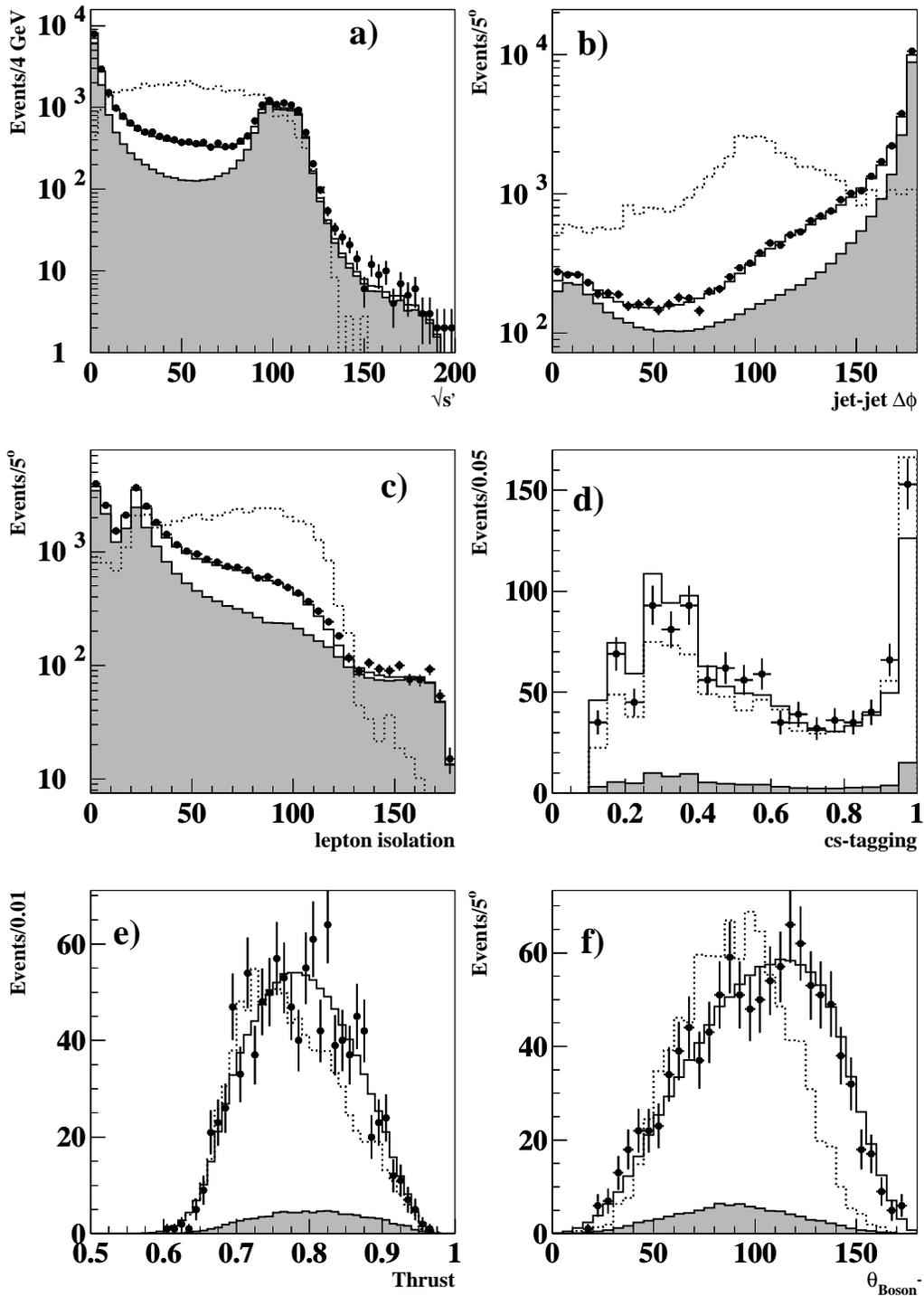,width=15.0cm}
\end{center}
\vspace{-1 cm}
\caption[]{ 
  Distributions of some of the variables used in the $\HCSTN$ analysis. 
  The effective centre-of-mass energy (a),
  $\varphi$ difference of hadronic jets (b), and the lepton isolation (c) 
  used in the anti-$\qq$ likelihood are shown after preselection. 
  The $\cs$-tagging variable (d), the thrust (e), 
  and the angle of the negatively charged boson (f) used in the anti-WW likelihood are shown at the final level. 
  Data are shown as filled circles, while
  the solid histogram contour shows the expected SM 
  background with contributions from $\ww$  (unfilled) and $\qq$ (shaded).
  The expected histogram for a
  75 GeV/$c^2$ charged Higgs boson signal is shown as a
  dotted histogram with
  arbitrary normalisation
  for comparison.}
\label{fig:varscstn}
\end{figure}

\begin{figure}[hbpt]
\begin{center}
  \epsfig{file= 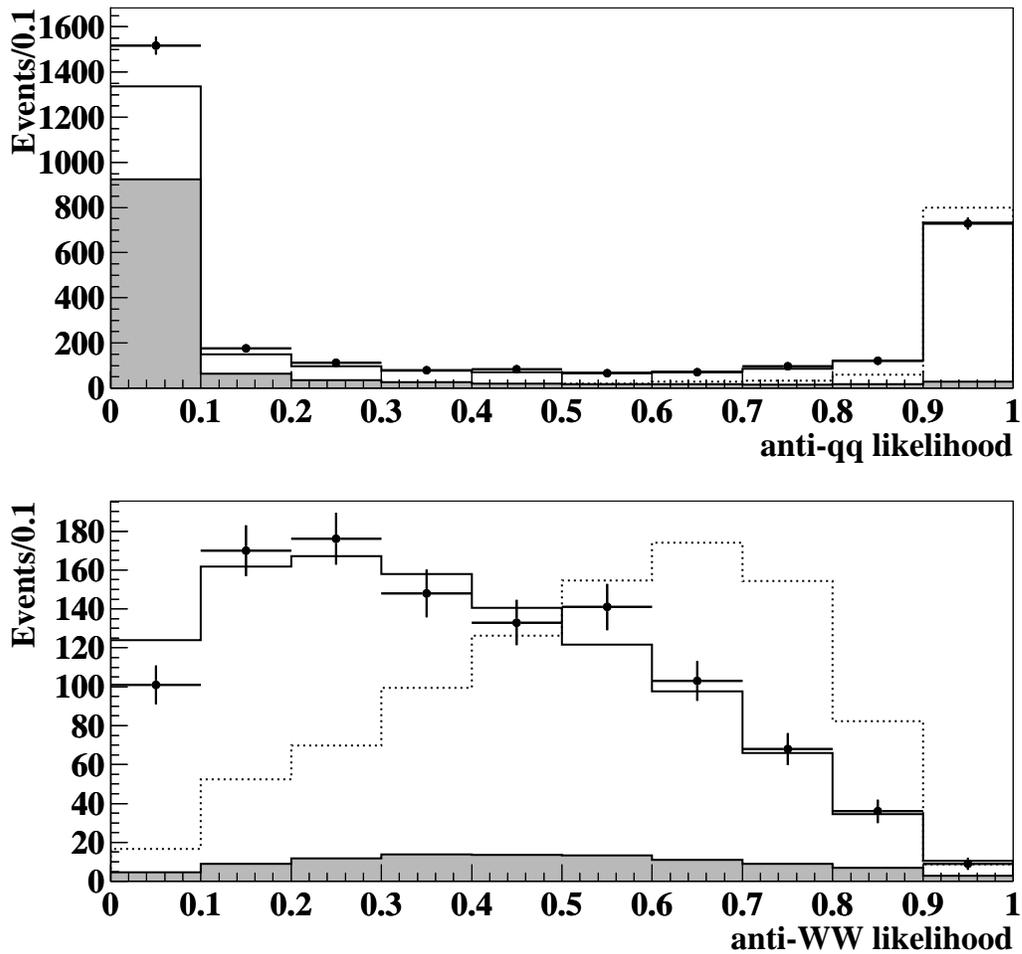,width=15.0cm}
\end{center}
\caption[]{Distributions of the anti-$\qq$  and anti-WW likelihoods
  for the $\HCSTN$ analysis 
  at $\sqs=$189--209~GeV\@. 
  The anti-$\qq$  likelihood is
  plotted after preselection and the anti-WW likelihood at the
  final level. 
  Data and SM background are indicated as in Fig. \ref{fig:varscstn}.
  The expected 
  distribution 
  for a 75 GeV/$c^2$ charged Higgs boson signal is shown as a
  dotted histogram with
  arbitrary normalisation.}
\label{fig:cstnlike}
\end{figure}

At this level of the selection there was still a very significant contribution of $\qq$
events. To reduce this background further a 
likelihood function was defined with eleven variables: the event thrust, the cosine of the
missing momentum, the angle in the transverse plane
 between the two hadronic jets, the reconstructed polar
angle of the negatively charged boson (with the charge defined
according to that of the $\tau$),
the angle between the $\tau$ jet and the parent boson's
momentum in the boson's rest-frame, the $\tau$ decay
channel, the total transverse momentum,
$\sqrt{s'} / \sqrt{s}$, the $\tau$ isolation, the DURHAM clustering distance $y_{3\rightarrow 2}$
when going from 
three to two jets
and the angle between the plane spanned by the 
two hadronic
jets and the $\tau$ candidate.
The latter angle took into account the fact that in most cases the $\qq$ background, produced when a radiated
gluon was confused with a $\tau$ jet, tended to have all three jets in the same plane, while 
for the signal the $\tau$ is more
or less uniformly distributed in space. 
Some of these variables are shown in Fig.~\ref{fig:varscstn} (a-c) and the likelihood 
is shown in Fig.~\ref{fig:cstnlike} (top). 
Events with an anti-$\qq$  likelihood lower than 0.5 were rejected.
The effects of the different
sets of cuts are shown in Table~\ref{tab:cstnsel} for the combined
$\sqs=$189--209~GeV sample. At preselection level the background from two-photon events
was slightly underestimated, due to the phase space cuts used in the generator. This is also
visible in figure~\ref{fig:cstnlike} (top).
Further cuts in the analysis are tighter than those in the generation and therefore 
the background estimation is not affected.

\begin{table} [ht!]
\begin{center}
\begin{tabular}{lrrrrr}
\hline
cut & \multicolumn{1}{c}{data}
    & \multicolumn{1}{c}{total bkg.}
    & \multicolumn{1}{c}{4-fermion}
    & \multicolumn{1}{c}{other bkg.}
    & \multicolumn{1}{c}{$\varepsilon_{75}$}\\ 
\hline
Preselection                & 31138 & 29803.1 & 9449.0 & 20354.1 & 95.8\% \\
Bulk $\qq$ rejection         &  6267 &  5899.7 & 3939.7 &  1960.0 & 84.9\% \\
$qq\tau\nu$ selection       &  3054 &  2814.5 & 1649.0 &  1165.4 & 66.1\% \\
anti-$\qq$  likelihood $>$ 0.5  &  1085 &  1081.7 &  985.8 &    95.9 & 57.5\% \\
\hline
\end{tabular}
\end{center}
\caption{The number of events selected in the data and expected from
Monte Carlo after the different cuts in the $\HCSTN$
analysis at $\sqs=$ 189--209~GeV\@. The efficiency in the last column corresponds to a
charged Higgs boson with a mass of 75 GeV$/c^2$.}
\label{tab:cstnsel}
\end{table}

\begin{figure}[hbpt]
\begin{center}
  \epsfig{file= 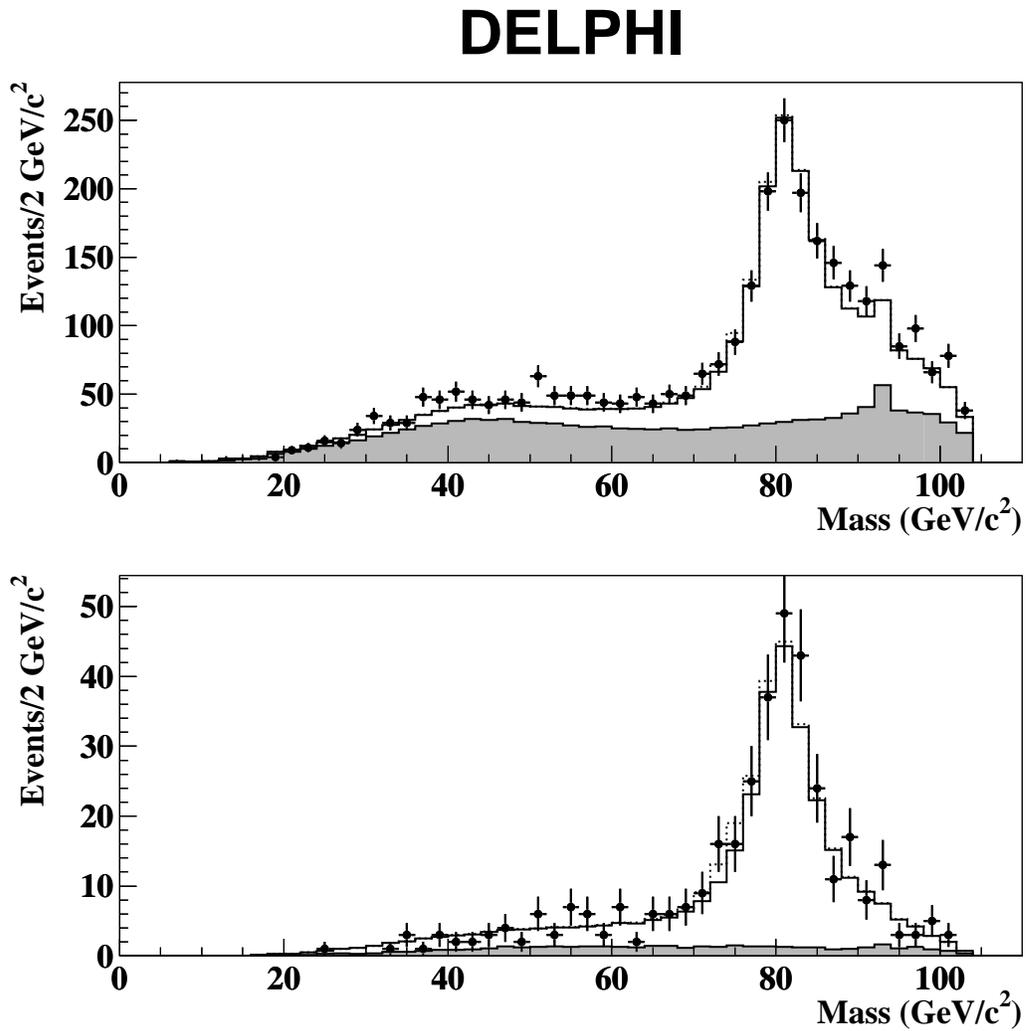,width=15.0cm}
\end{center}
\caption[]{Reconstructed mass distribution for events selected
  in the $\HCSTN$ analysis
  at $\sqs=$189--209~GeV at preselection (top) and at the final selection level (bottom), 
  with an additional
   cut on the anti-WW likelihood 
   ${\cal L}_{\mathrm{WW}} > 0.5$ . 
  Data and SM background are indicated as in Fig. \ref{fig:varscstn}.
   The expected distribution in the presence of an 
   H$^+$H$^-$ signal, with $\mhp = 75~\mbox{\rm GeV}/c^2$
   and leptonic branching ratio of 50\%, is also shown for 
   comparison (dotted).
}
\label{fig:cstnmass}
\end{figure}

At this stage, most of the remaining background was $\ww$ decaying to
$\qq\tn$, whose topology is equivalent  to that of the signal. 
Further background rejection was possible, however, using
the $\tau$ polarisation and the output of the jet flavour algorithm. 
Another likelihood function was therefore defined
using these
two variables
and
some of the variables used in 
the previous anti-$\qq$  likelihood since 
these
also improved the $\ww$  
rejection.
The additional variables were
the thrust, angle in the transverse plane between the two hadronic jets, the reconstructed polar
angle of the negatively charged boson, the angle between the $\tau$
momentum and its parent boson's momentum in the boson's rest-frame and the $\tau$ isolation angle.
Some of these variables are shown in Fig.~\ref{fig:varscstn} (d-f) and the result of the likelihood 
is shown in Fig.~\ref{fig:cstnlike} (bottom). No cut was imposed on this function, but it was used in the
limit estimation as described below. However, Fig.~\ref{fig:cstnmass}, shows the mass distribution
after a cut on ${\cal L}_{\mathrm{WW}} > 0.5$ to visually enhance the mass distribution of the events whose variables are closer
to those expected for the charged Higgs signal.

\subsection{Channels including a $\wa$ decay}

If at least one of the Higgs bosons decays to a $\wa$ pair, there are several possible
topologies depending on the different boson decays.  The W  can 
decay leptonically or hadronically, and the
number of jets strongly depends on the A mass and on the 
boson
boosts. The search was restricted to A
masses above 12 GeV$/c^2$, where it decays predominantly to $\bb$ and an inclusive
search was performed. Events with jets with b quark content 
were
searched for in two
topologies:
\begin{itemize}
\item events with a $\tau$, missing energy and at least two hadronic jets
\item events with no missing energy and at least four hadronic jets
\end{itemize}
In this way  
most of the possible decay chains for the $\HWATN$ (first topology) and 
$\HWAWA$ (second) 
were covered. The decay to $\HWACS$ 
was neglected because its
contribution is small. Its branching ratio 
is usually very small and only reaches
a maximum of about $17\%$ in a small region of the parameter space. In this
region the branching ratio for $\HWATN$ is about twice as large, with a smaller
background. The branching ratio for $\HWAWA$ is about $30\%$, giving a signal 
 almost indistinguishable from $\HWACS$. 

The analysis designed by DELPHI for technipion search within Technicolor models~\cite{delphitc}
was well suited also for these topologies and had a good
performance in this search. It was therefore adopted here. A brief description of that analysis is outlined here.

\subsubsection{Semileptonic final states.}

Since the topology searched for in the semileptonic case is very close to 
the corresponding channel in 
$\ww$ production, a selection similar to that used on $\ww$ cross-section and 
decay branching ratio measurements~\cite{delphiww}
was applied at the first step. However, variables strongly correlated 
with the boson mass were not used, making the analysis efficient for a wide 
range of masses. 

Loose initial cuts, requiring at least seven charged particles,
transverse energy greater than $0.25 \sqrt{s}$, less than 30 GeV within a 
$30^\circ$ cone around the beam-axis, and the polar angle of the missing momentum 
fulfilling $|\cos{\theta_{miss}}|<0.985$, were used to remove a large fraction
of the leptonic, $q\bar q(\gamma)$ and events from two-photon interactions.

Then, an isolated $\tau$ candidate had to be found, to reduce the background from
$\ww$ leptonic decays not involving $\tau$ particles. 
The isolation criterion was defined 
in terms of the product $p \cdot \theta_{iso}$, where 
$p$ is the $\tau$ jet momentum and $\theta_{iso}$ is the isolation angle between 
the lepton and the nearest charged particle with momentum greater than 
1~GeV/c. 

To reject non-$\ww$ events, a neural 
network (NN) with the following variables was used: 
the $\tau$ jet momentum,  $\tau$  jet isolation angle,
 cosine of the polar angle of the missing momentum, transverse momentum,
 thrust, angle between the lepton and the hadronic system, 
 the acoplanarity and acollinearity of the
hadronic jets and $\sqrt{s'/s}$ .
Events were accepted if the NN output was above 0.6. 
In this way most of the non-$\ww$ background is rejected. 

\begin{figure}[hbpt]
\begin{center}
  \epsfig{file= 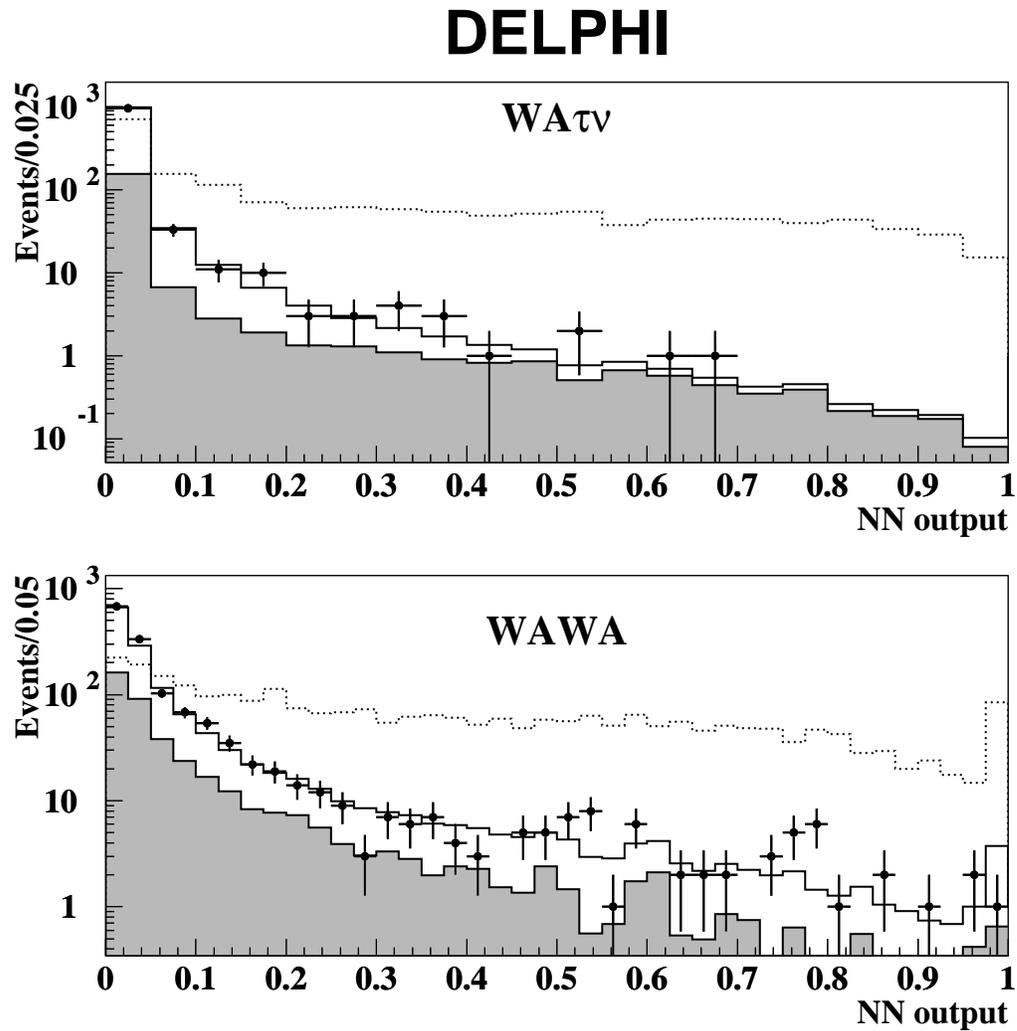,width=15.0cm}
\end{center}
\caption[]{Distribution of the output of the final
discriminating neural network 
for events selected in the
$\HWATN$ (top) and 
  $\HWAWA$ (bottom) analyses after preselection and a cut of 0.01 in the variable plotted, for centre-of-mass energies 
  between 189 and 209~GeV\@. 
  The data and the simulated SM background are indicated as in previous figures. 
   The expected distribution in the presence of an 
   H$^+$H$^-$ signal, with $\mhp = 80~\mbox{\rm GeV}/c^2$
   and $\ma = 30$ GeV/$c^2$, is also shown in arbitrary normalisation for 
   comparison (dotted).
}
\label{fig:wann}
\end{figure}

The second step exploits the specific properties of
the signal, such as the presence of b-quarks or the production angle, to
distinguish it from the W pairs.
This is done using another neural network which uses four input variables: 
the b-tagging variables~\cite{btag} of the two hadronic jets, 
the signed cosine of the polar angle of the 
boson\footnote{ The 
sign is defined by the charge of the $\tau$, 
and the production polar angle $\theta_{prod}$ is the one obtained from the 5C
fit.}  and $|\cos{\theta_{miss}}|$.
The distribution of the output of this neural network for
signal and background is shown in Fig.~\ref{fig:wann} (top), at the final level of the analysis 
with an additional cut at 0.01 for better presentation removing a large peak at 0. 
This variable was not used for selection, but just as additional
discriminant information for the confidence level estimation.

\begin{figure}[hbpt]
\begin{center}
  \epsfig{file= 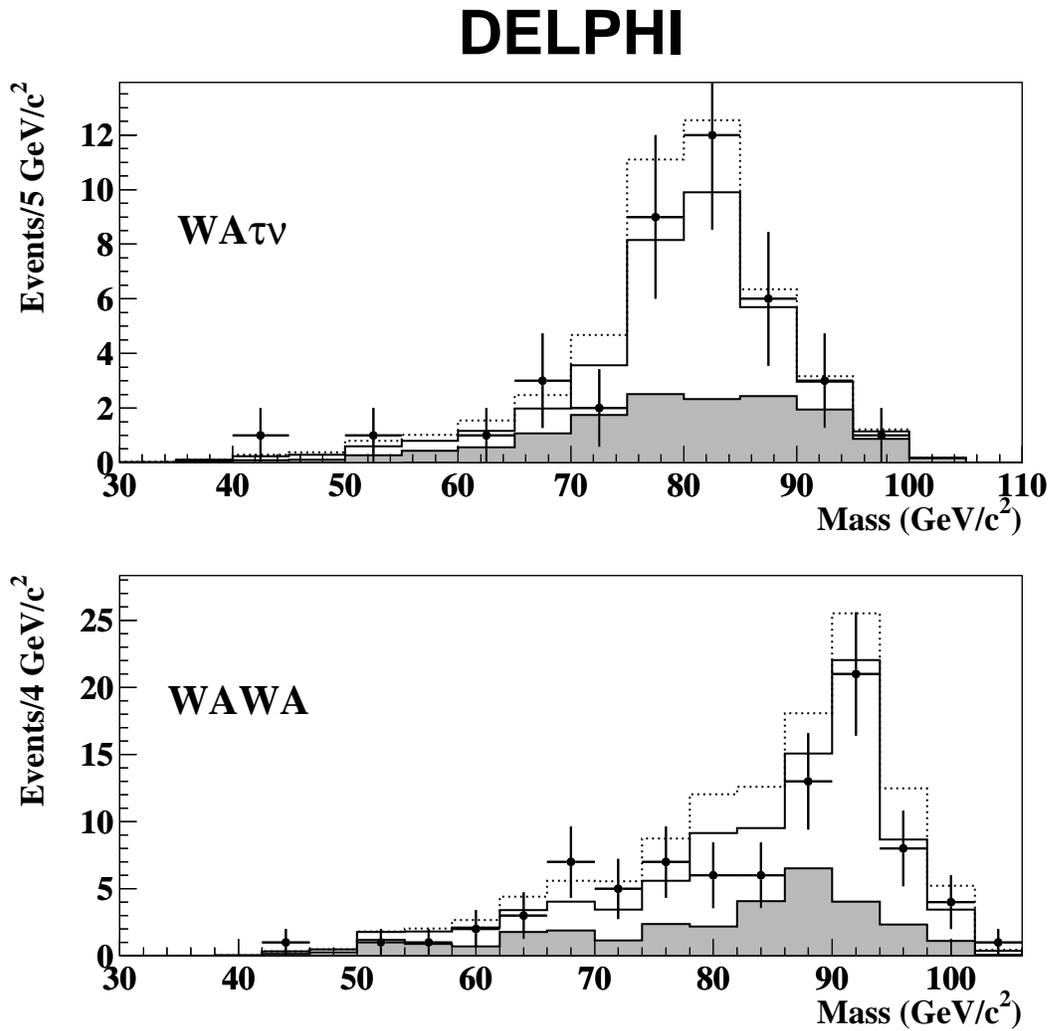,width=15.0cm}
\end{center}
\caption[]{Reconstructed mass distribution 
for events selected in the
$\HWATN$ (top) and 
  $\HWAWA$ (bottom) analyses by a
  cut on the neural network output of 0.1 and 0.3,
  respectively, for centre-of-mass energies 
  between 189 and 209~GeV\@. 
  The data and the simulated SM background are indicated as in previous figures. 
   The expected distribution in the presence of an 
   H$^+$H$^-$ signal, with $\mhp = 80~\mbox{\rm GeV}/c^2$
   and $\ma = 30$ GeV/$c^2$, is also shown for 
   comparison (dotted).
}
\label{fig:wamass}
\end{figure}

The effects of the different sets of cuts are shown in
Table~\ref{tab:watnsel}  for the combined $\sqs=$189--209~GeV sample. 
Fig.~\ref{fig:wamass} (top) shows the reconstructed mass, using a 5C fit, of the selected candidates.

\begin{table} [ht!]
\begin{center}
\vspace{1ex}
\begin{tabular}{lrrrrr}
\hline
cut & \multicolumn{1}{c}{data}
    & \multicolumn{1}{c}{total bkg.}
    & \multicolumn{1}{c}{4-fermion}
    & \multicolumn{1}{c}{other bkg.}
    & \multicolumn{1}{c}{$\varepsilon_{80/30}$} \\
\hline
Hadronic preselection            &  28380 & 28274.8 & 3925.9 &    24348.9 & 88.9\% \\
$\qq \tau \nu$ selection         &   1043 &  1061.9 &  884.5 &      177.4 & 44.0\% \\
NN output $> 0.1$                &     39 &    36.8 &   22.2 &       14.6 & 22.6\% \\
NN output $> 0.2$                &     18 &    17.8 &    7.9 &        9.9 & 17.9\% \\
NN output $> 0.3$                &     12 &    11.0 &    3.7 &        7.3 & 14.9\% \\
\hline
\end{tabular}
\end{center}
\caption[]{The total number of events observed and expected backgrounds 
in the $\HWATN$ channel after the different cuts used in the analysis at $\sqs=$ 189--209~GeV\@. 
The last column shows the efficiencies for charged Higgs boson signals 
with $\mhp = 80$~GeV/$c^2$ and $\ma$= 30~GeV/$c^2$.}
\label{tab:watnsel}
\end{table}

\subsubsection{Hadronic final states. }
The $\HWAWA$ analysis started with 
the four-jet preselection used in the search for neutral Higgs bosons~\cite{pap97},
which aimed to eliminate the $q\bar{q}(\gamma)$ and events from two-photon interactions and to
reduce the QCD and $\zo\gamma^*$ background.
The $q\bar{q}(\gamma)$ and 4-fermion backgrounds remaining after the 
preselection had to be reduced further. For this purpose different shape and 
${\mathrm b}$-tagging variables have been investigated. 
Finally, 12 variables were selected for this analysis and the
final discriminant variable was defined as the output of a
neural network.
There were two ${\mathrm b}$-tagging variables intended 
to reduce the $\ww$ background: one of them ($x_b$) was computed as 
the sum of the two highest jet ${\mathrm b}$-tagging variables, 
and the other was the sum of the four jet ${\mathrm b}$-tagging variables.
Seven shape variables were used to reduce the \qqg\ contamination. 
They  were 
the sum of the second and fourth Fox-Wolfram moments,
the product of the minimum jet energy and the minimum opening angle between
any two jets,
the event thrust,
the sum of  the four lowest angles between any pair of jets in the event,
the minimal di-jet mass,
and the minimal $y_{cut}$ values
for which the event was clustered into 4 jets ($y_{4\to3}$)
and into 5 jets ($y_{5\to4}$). Finally, three more variables 
took into account the two-boson event topology. To define them the event 
was forced into four jets, a 5C fit, requiring conservation 
of energy and momentum and equal masses of opposite jet pairs, was applied to 
all possible jet pairings, and the pairing giving the smallest 
value of the fit $\chi^2$ was selected. 
The variables then included in the neural network were 
the smallest $\chi^2$,
the production angle of the jet pair, 
and the angle between the planes defined by the two jet pairs. 
The distribution of the output of this neural network for
signal and background is shown in Fig.~\ref{fig:wann} (bottom), at the final level of the analysis 
with an additional cut at 0.01 for better presentation removing a large peak at 0. 
This variable was not used for selection, but just as additional
discriminant information for the confidence level estimation.

The effects of the different sets of cuts are shown in
Table~\ref{tab:wawasel} for the combined $\sqs=$189--209~GeV sample. 
Fig.~\ref{fig:wamass} (bottom) shows the reconstructed mass, using a 5C fit, of the selected candidates.

\begin{table} [ht!]
\begin{center}
\vspace{1ex}
\begin{tabular}{lrrrrr}
\hline
cut & \multicolumn{1}{c}{data}
    & \multicolumn{1}{c}{total bkg.}
    & \multicolumn{1}{c}{$\qq g$}
    & \multicolumn{1}{c}{4-fermion}
    & \multicolumn{1}{c}{$\varepsilon_{80/30}$} \\
\hline
         preselection            &   6592 &   6520.1 & 2004.9 &    4515.2 & 67.9\% \\
NN output $> 0.1$                &    253 &    252.5 &   87.2 &     165.3 & 46.1\% \\
NN output $> 0.3$                &     86 &     78.9 &   25.2 &      53.7 & 28.3\% \\
\hline
\end{tabular}
\end{center}
\caption[]{The total number of events observed and expected backgrounds 
in the $\HWAWA$ channel after the different cuts used in the analysis at $\sqs=$ 189--209~GeV\@. 
The last column shows the efficiencies for charged Higgs boson signals 
with $\mhp = 80$~GeV/$c^2$ and $\ma$= 30~GeV/$c^2$.}
\label{tab:wawasel}
\end{table}

\section{Systematic errors}

Uncertainties in the expected background rates and in the signal efficiency
were accounted for at each centre-of-mass energy and separately for the S6 period. 
Small contributions to the background rate uncertainties, of the order of $0.6 \%$, are due
to uncertainties in the luminosity measurement and in the theoretical
cross-section estimates  
for the simulated data samples.
The
systematic 
error
estimation 
for 
the background follows closely
the treatment 
in the DELPHI  $\ww$  cross-section measurement
\cite{delphiww}. 

The largest part of the background and signal efficiency
uncertainties in the $\HTT$ channel is due to the limited simulation
statistics available. The typical contribution was $8\%$ and $1.5\%$, respectively.
Several additional sources of systematic uncertainties were investigated. In particular,
the track reconstruction efficiency, the $\tau$ identification and the behaviour
of different variables were
studied. 

The systematic errors induced by the track reconstruction and $\tau$ identification 
were checked by a comparison with independent samples of di-lepton or two-photon
leptonic events
of simulation and real data, taken with the same detector conditions both at high energy and at the $\zo$  
resonance. These
samples were selected by kinematic cuts, with only very loose  particle identification requirements, 
which were found to be
uncorrelated to those used in this analysis.
The lepton identification efficiency estimate from data and  simulation 
was found to agree within the statistical errors  (about 1\%). 
The same leptonic samples were used to check the track reconstruction efficiency 
of isolated particles,
showing an
agreement at
the 1\% level.
The modelling of the preselection
variables agrees within statistical errors with the data.  
The momentum and electromagnetic energy scales and resolutions were investigated using radiative di-lepton events,
$\mu^+\mu^-\gamma$ or $\mathrm{e^+e^-}\gamma$, from data and 
simulation. For these events, the momenta of the particles can be calculated with very good precision from kinematical
constraints, independently of the direct measurement on the tracking detectors or calorimeters, allowing comparisons.
In all cases, data and simulation agreed to better than the statistical precision,
with a negligible overall influence both on the signal efficiency and on the background rates.
Additional systematic effects were estimated by comparing the data collected
at 
the $\zo$
peak during the period 
when sector 6 of the TPC was not functioning
with
simulation samples produced with the same detector 
conditions. This did not indicate any significant increase in the systematic errors, 
compared to those quoted above. 
The total
systematic error on the signal efficiency was 2\% and the total
relative systematic error on the background rate was 10\%.


In the $\HCSCS$ analysis, the total uncertainty of the $\qq g g$ background estimate at the
four-jet preselection level was dominated~\cite{delphiww} by the hadronisation model
and imperfections in the generator model. 
Based on a comparison of
three models provided by the generators
 JETSET 7.4~\cite{pythia}, HERWIG~\cite{herwig} and ARIADNE~\cite{ariadne}, the total
uncertainty of the $\qq g g$ event rate 
was
estimated to be of the order of 5\%.

Another uncertainty 
in the
four-fermion background (mainly $\ww$),
is due to the uncertainties in the luminosity
measurement and in the cross-section estimate. The precision of the
Standard Model prediction for the $\ww$  production cross-section estimate
depends on the centre-of-mass energy and has been estimated to be of
the order of 1\%.   
An additional systematic error on the background rate arose from 
the preselection efficiency precision. The detailed study made in~\cite{delphiww}
could also be applied to this analysis, leading to a total uncertainty of 0.6\%.  
The main contribution to this uncertainty 
is also
the hadronisation model, 
with
smaller contributions from the detector
simulations. 
Combining these uncertainties the estimated precision of
the four-fermion background rate at the preselection level was 1.3\%.

Further systematic effects could have been introduced in the analysis
when applying the relative mass difference cut and the likelihood
background rejections. Any  differences in the shapes of
these variables between the real and simulated data would affect the
efficiency of the cuts. 
Comparisons 
were made
at early selection levels in order to keep the event rates 
high, enabling large statistics for the comparisons and 
keeping the signal-to-background rate adequately small so that a possible 
signal in the data would not affect the distributions 
significantly. The uncertainty on the background rate due to the
relative mass difference cut was estimated to be 1\%. The effect of
potential systematic effects of the shapes of the likelihood variable
distributions was studied by changing the variable shapes in the
simulation by reweighting simulated events. The reweighted events were
propagated through the analysis and the effect on the cut efficiencies
was studied. The uncertainty of the anti-$\qq$  likelihood and anti-WW
cuts were estimated to be 2.3\% and 0.7\%. 
Uncertainties in
the final discriminating likelihood shape, which would 
affect
the signal likelihood of the data events, 
were
also taken into
account. 
A change
in the likelihood shape would 
influence the
likelihood ratio in the exclusion limit
calculation. This effect was taken into account by increasing the
background rate uncertainty by an additional 2\%.

Combination of all background uncertainties leads to a total
uncertainty of 4\% 
in
the background normalisation. The uncertainty of
the signal efficiency was estimated to be 2.5\% with a 1\%
contribution from beam energy, hadronisation model etc., a 1.2\%
contribution from limited 
simulation
statistics and a 2\% contribution from
the cuts and likelihoods.

In the  $\HCSTN$ channel, the contribution to the systematic error from the
uncertainties in the $\qq$  and $\ww$  total normalisation was estimated in 
a similar way to be 0.4\% and 0.9\%, 
respectively. 
The isolated lepton
identification efficiency was estimated with the same di-lepton samples used for the $\HTT$ channel, 
with a 
contribution  of 1\% both to the signal and background systematics.
The uncertainties of the selection variables
were
estimated by comparing the shapes of the variable
distributions in data and 
simulation at the preselection level.
All variables agreed within statistical errors. 
Nevertheless, the potential error was estimated conservatively from the observed difference
between real data and simulation
when any particular cut was varied within the resolution of the corresponding variable.
Combining these errors, a
total uncertainty of $2.4$\% 
was
estimated for the background
rate and $0.3\%$ in the signal efficiency. For the likelihood functions, the 
reweighting procedure
described for 
$\HCSCS$ was followed, estimating the total contribution 
to $7.6\%$ for the
background and $3.2\%$ for the signal.

For the $\HWATN$ and the $\HWAWA$ channels, a similar procedure was followed,
with an additional contribution from the b-tagging and with the
difference that the $\ww$  is not the dominant background (described in
detail in~\cite{delphitc}). 
The total systematic errors on the signal efficiency for the $\HWAWA$ and $\HWATN$ were 5\% and
2\% respectively. The relative errors on the background were 11\% and 10\%.

\section{Results}

The number of  data and background events and the estimated
efficiencies in each of the analysis channels for different H$^{\pm}$ masses
are summarised in Tables~\ref{tab:tabeff} and~\ref{tab:tabeffwa}.
The quoted errors include both statistic and systematic errors, added in quadrature. 


\begin{table}[hb!]
\begin{center}
\begin{tabular}{cccrrr@{$\pm$}rr@{$\pm$}rr@{$\pm$}r}
\hline
& Chan. 
& $\sqrt{s}$ (GeV) 
& \multicolumn{1}{c}{lum.}
& \multicolumn{1}{c}{data}
& \multicolumn{2}{c}{total bkg.}
& \multicolumn{2}{c}{$\varepsilon_{75}$} 
& \multicolumn{2}{c}{$\varepsilon_{80}$} \\
\hline 
\hline
& $\HTT$ & 189       & 153.8  & 14 & 17.8 & 1.4 & 35.2 & 1.5\% & 35.7 & 1.5\% \\
& $\HTT$ & 192       &  24.5  &  3 &  2.9 & 0.2 & 33.6 & 1.5\% & 37.0 & 1.5\% \\
& $\HTT$ & 196       &  72.4  & 10 &  9.1 & 0.7 & 33.6 & 1.5\% & 37.0 & 1.5\% \\
& $\HTT$ & 200       &  81.8  & 10 &  9.7 & 0.8 & 32.3 & 1.5\% & 35.5 & 1.5\% \\
& $\HTT$ & 202       &  39.4  &  2 &  4.7 & 0.4 & 32.3 & 1.5\% & 35.5 & 1.5\% \\
& $\HTT$ & 205       &  69.1  & 10 &  8.5 & 0.6 & 32.2 & 1.5\% & 33.4 & 1.5\% \\
& $\HTT$ & 206.6     &  79.8  &  5 & 10.1 & 0.8 & 32.2 & 1.5\% & 33.4 & 1.5\% \\
& $\HTT$ & 206.3(S6) &  50.0  &  5 &  6.1 & 0.5 & 31.7 & 1.5\% & 35.7 & 1.5\% \\
\hline
& $\HCSCS$ & 189        &    158.0   &  565 & 554.9 & 22.2  & 52.1 & 1.3\% & 52.6 & 1.3\% \\
& $\HCSCS$ & 192        &     25.9   &   90 &  93.1 &  3.7  & 54.6 & 1.4\% & 54.1 & 1.4\% \\  
& $\HCSCS$ & 196        &     76.9   &  284 & 279.7 & 11.2  & 54.6 & 1.4\% & 54.1 & 1.4\% \\
& $\HCSCS$ & 200        &     84.3   &  299 & 300.6 & 12.2  & 53.1 & 1.3\% & 53.9 & 1.3\% \\
& $\HCSCS$ & 202        &     41.1   &  147 & 136.5 &  5.5  & 53.1 & 1.3\% & 53.9 & 1.3\% \\
& $\HCSCS$ & 205        &     75.6   &  270 & 264.5 & 10.6  & 51.5 & 1.3\% & 53.6 & 1.3\% \\
& $\HCSCS$ & 206.6      &     87.8   &  291 & 288.3 & 11.5  & 52.1 & 1.3\% & 53.5 & 1.3\% \\
& $\HCSCS$ & 206.3 (S6) &     60.8   &  168 & 196.9 &  7.9  & 51.5 & 1.3\% & 53.6 & 1.3\% \\
\hline
& $\HCSTN$ &  189         & 153.8 &  296 & 285.8 & 22.9 & 57.5 & 2.7\% & 57.1 & 2.7\% \\
& $\HCSTN$ &  192         &  24.5 &   56 &  47.5 &  3.8 & 57.6 & 2.7\% & 56.5 & 2.7\% \\
& $\HCSTN$ &  196         &  72.4 &  147 & 143.8 & 11.5 & 57.6 & 2.7\% & 56.5 & 2.7\% \\
& $\HCSTN$ &  200         &  81.8 &  158 & 154.6 & 12.4 & 57.4 & 2.7\% & 57.3 & 2.7\% \\
& $\HCSTN$ &  202         &  39.4 &   71 &  75.7 &  6.1 & 57.4 & 2.7\% & 57.3 & 2.7\% \\
& $\HCSTN$ &  205         &  69.1 &  130 & 129.5 & 10.4 & 57.2 & 2.7\% & 55.5 & 2.6\% \\
& $\HCSTN$ &  206.6       &  79.8 &  139 & 150.4 & 12.0 & 57.2 & 2.7\% & 55.5 & 2.6\% \\
& $\HCSTN$ &  206.3(S6)   &  50.0 &   88 &  94.4 &  7.6 & 57.7 & 2.7\% & 55.9 & 2.6\% \\
\hline
\end{tabular}
\end{center}
\caption[]{Integrated luminosity, observed number of events, 
expected number of background events and signal efficiency 
(for 75~GeV/$c^2$ and 80~GeV/$c^2$ charged Higgs boson masses) for 
different centre-of-mass energies for the channels not involving $\wa$ decays.} 

\label{tab:tabeff}

\end{table}

\begin{table}[h!]
\begin{center}
\begin{tabular}{cccrrr@{$\pm$}rr@{$\pm$}rr@{$\pm$}r}
\hline
& Chan. 
& $\sqrt{s}$ (GeV) 
& \multicolumn{1}{c}{lum.}
& \multicolumn{1}{c}{data}
& \multicolumn{2}{c}{total bkg.}
& \multicolumn{2}{c}{$\varepsilon_{80}$} 
& \multicolumn{2}{c}{$\varepsilon_{90}$} \\
\hline 
\hline
& $\HWATN$ & 189       & 153.8  & 12 & 11.4 & 0.7 & 20.5 & 2.2\% & 10.2 & 2.1\% \\
& $\HWATN$ & 192       &  24.5  &  3 &  1.6 & 0.1 & 20.1 & 2.2\% & 11.4 & 2.1\% \\
& $\HWATN$ & 196       &  72.4  &  2 &  4.7 & 0.3 & 20.1 & 2.2\% & 11.4 & 2.1\% \\
& $\HWATN$ & 200       &  81.8  &  4 &  4.9 & 0.3 & 21.0 & 2.2\% & 13.7 & 2.1\% \\
& $\HWATN$ & 202       &  39.4  &  4 &  2.5 & 0.2 & 21.0 & 2.2\% & 13.7 & 2.1\% \\
& $\HWATN$ & 205       &  69.1  &  4 &  4.1 & 0.2 & 21.3 & 2.2\% & 15.5 & 2.2\% \\
& $\HWATN$ & 206.6     &  79.8  &  6 &  4.6 & 0.3 & 21.3 & 2.2\% & 15.5 & 2.2\% \\
& $\HWATN$ & 206.3(S6) &  50.0  &  4 &  3.0 & 0.2 & 21.3 & 2.2\% & 15.5 & 2.2\% \\
\hline
& $\HWAWA$ & 189       & 158.0  & 81 & 79.7 & 7.9 & 35.6 & 5.1\% & 39.4 & 5.1\% \\
& $\HWAWA$ & 192       &  25.9  & 16 & 13.0 & 1.3 & 35.6 & 5.1\% & 39.4 & 5.1\% \\
& $\HWAWA$ & 196       &  76.9  & 37 & 35.3 & 3.5 & 35.6 & 5.1\% & 39.4 & 5.1\% \\
& $\HWAWA$ & 200       &  84.3  & 36 & 35.6 & 3.6 & 35.5 & 5.1\% & 39.3 & 5.1\% \\
& $\HWAWA$ & 202       &  41.1  & 16 & 17.7 & 1.8 & 35.5 & 5.1\% & 39.3 & 5.1\% \\
& $\HWAWA$ & 205       &  75.6  & 24 & 24.7 & 2.5 & 37.8 & 5.1\% & 34.5 & 5.1\% \\
& $\HWAWA$ & 206.6     &  87.8  & 30 & 28.3 & 2.8 & 37.8 & 5.1\% & 34.5 & 5.1\% \\
& $\HWAWA$ & 206.3º(S6) &  60.8  & 13 & 18.2 & 2.8 & 37.8 & 5.1\% & 34.5 & 5.1\% \\
\hline
\end{tabular}
\end{center}
\caption[]{Integrated luminosity, observed number of events, 
expected number of background events and signal efficiency 
(for 80~GeV/$c^2$ and 90~GeV/$c^2$ charged Higgs boson masses, and $\ma$=12~GeV/$c^2$) for 
different centre-of-mass energies for the channels involving $\wa$ decays.} 
\label{tab:tabeffwa}
\end{table}

\subsection{Determination of the mass limit}

No significant signal-like excess of events compared to the expected
backgrounds was observed in any of the five final states
investigated.  
Confidence levels were calculated using a modified frequentist technique, based on the
extended maximum likelihood ratio~\cite{alex,higgs}.
From these confidence levels, lower limits on the charged Higgs boson mass were
derived at 95\% confidence level in two scenarios. 
In the first scenario
it was assumed that the charged
Higgs boson decayed exclusively to either $\tn$ or $\cs$, corresponding to type II models.
 The limits were extracted
as a function of the leptonic Higgs
decay branching ratio BR($H^-\rightarrow \tn$). 
In the second scenario, corresponding to type I models,
the $\wa$ decay was permitted if kinematically accessible
and limits were computed for different values of $\ma$ as a function of $\tan\beta$ or 
for different values of $\tan\beta$ as a function of $\ma$. The
branching ratios were calculated according to~\cite{WAteo} as functions of $\tan\beta$ and 
the neutral pseudo-scalar and charged Higgs masses.

The background and signal probability density functions of one or two
discriminating variables in each channel were used. The data samples
collected at the different centre-of-mass energies were treated 
as independent channels. When there was a significant overlap between two channels, the one providing less sensitivity
was ignored to avoid double counting. In the $\HCSCS$ and $\HCSTN$
 channels the two discriminating variables were the
reconstructed Higgs boson mass and the anti-WW likelihood. In the $\HWAWA$ and $\HWATN$ the likelihood was
replaced by the final neural network output.
In the $\HTT$
channel only the final background discrimination likelihood was used since
mass reconstruction was not possible. The distributions of the
discriminating variable for signal events, obtained from the simulation
at different H$^\pm$ mass values for each $\sqrt{s}$, were
interpolated for intermediate mass values. 

The estimated uncertainties on
background and signal were taken into account in the limit derivation by a Gaussian
smearing around the central values of the number of expected events.

\begin{figure}[hbpt]
\begin{center}
\epsfig{file=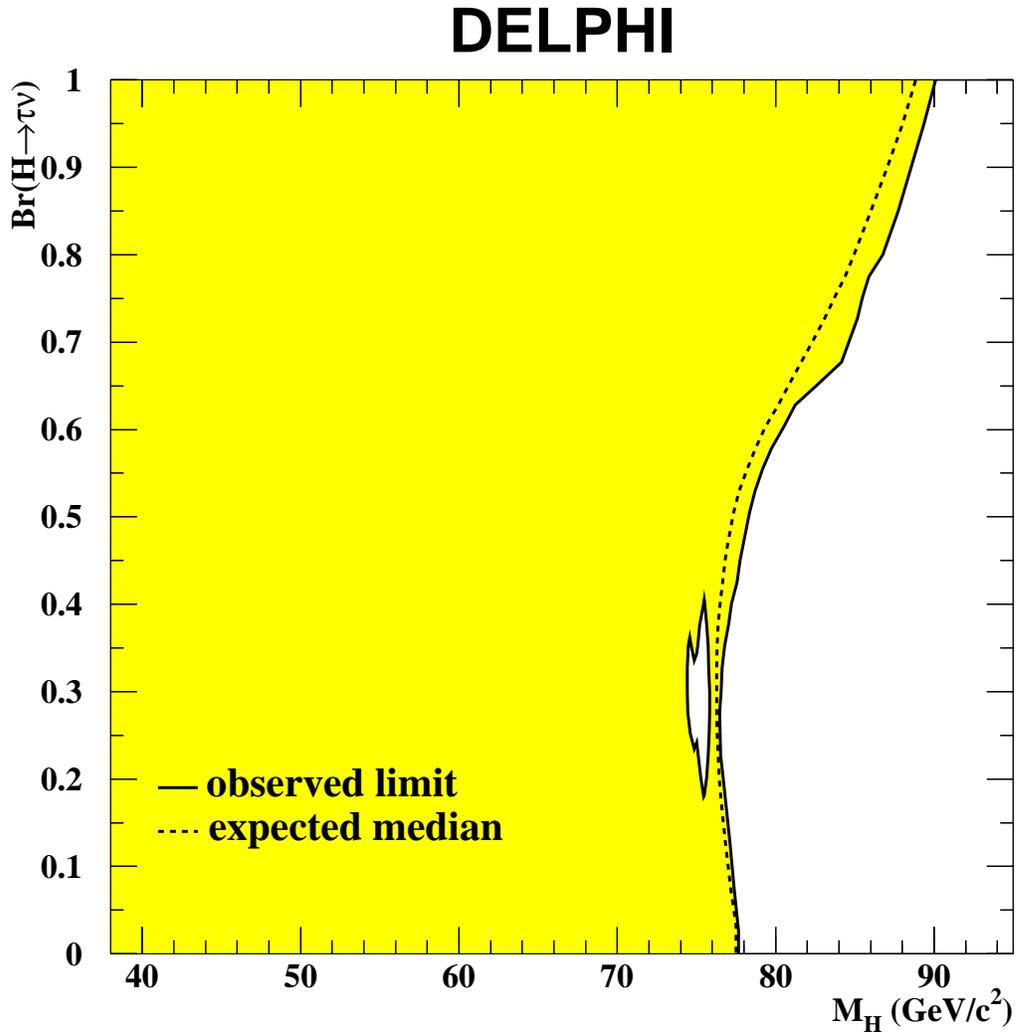,width=15.0cm}
\end{center}
\caption[]{
     The observed and expected exclusion regions at 95\% confidence level in the
     plane of BR($H^- \rightarrow \tn$) vs. $\mhp$.
     These limits were obtained from a combination of the search results in the
      $\HTT$, $\HCSTN$ and $\HCSCS$ channels at  $\sqrt{s}=$ 189--209~GeV, under
     the assumption that the $\wa$ decay is forbidden.
   }
\label{fig:limit}
\end{figure}

\begin{figure}[hbpt]
\begin{center}
\epsfig{file=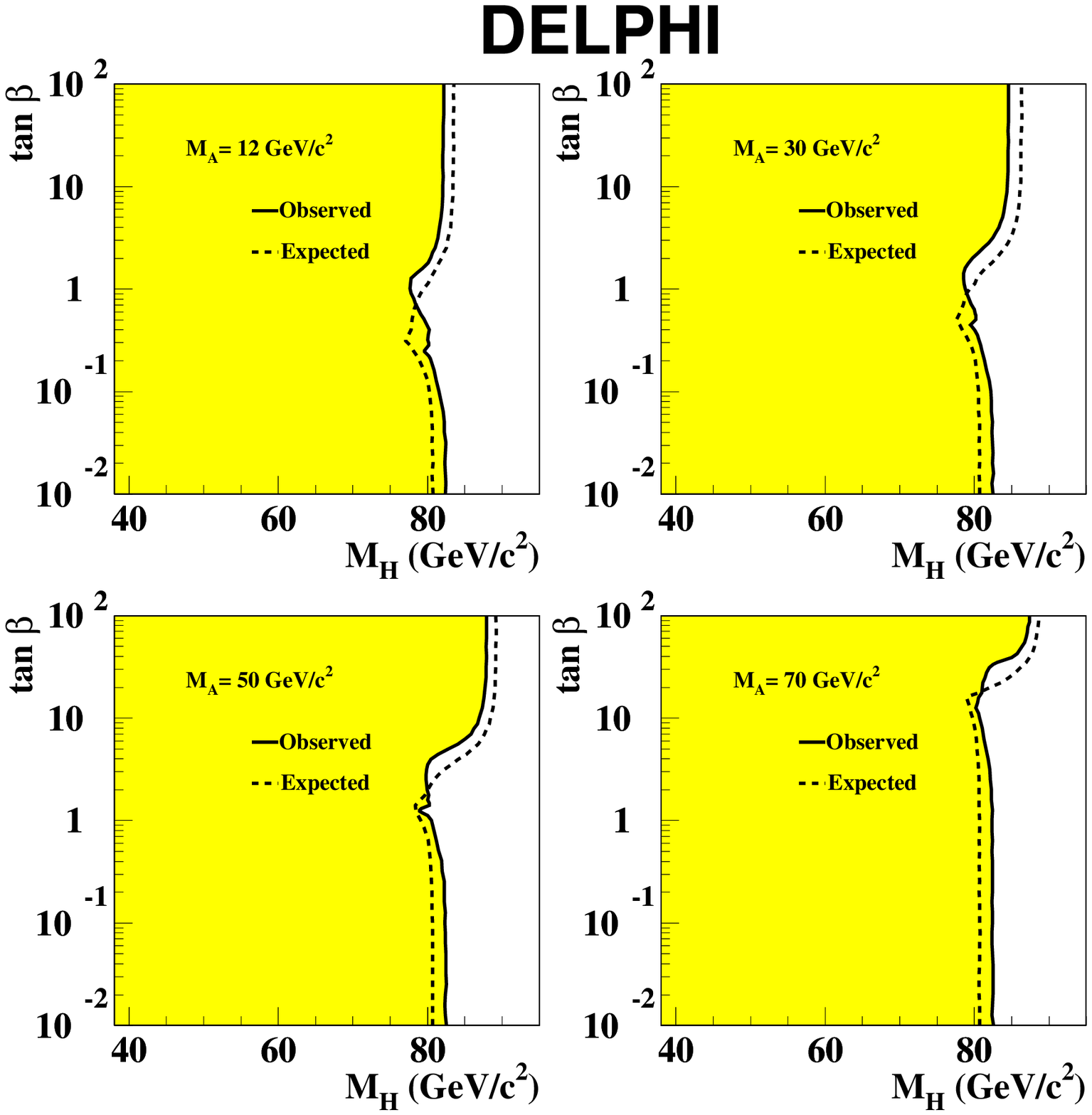,width=15.0cm}
\end{center}
\caption[]{
  The observed and expected exclusion regions at 95\% confidence level in the
  plane of  $\tan\beta$  vs. $\mhp$ in the framework
     of type I Two Higgs Doublet Models. These limits were
  obtained from a combination of the search results in all studied channels, with or without $\wa$ decays,
   at $\sqrt{s}=$ 189--209~GeV, for different A masses.} 
\label{fig:limitwa}
\end{figure}

\begin{figure}[hbpt]
\begin{center}
\epsfig{file=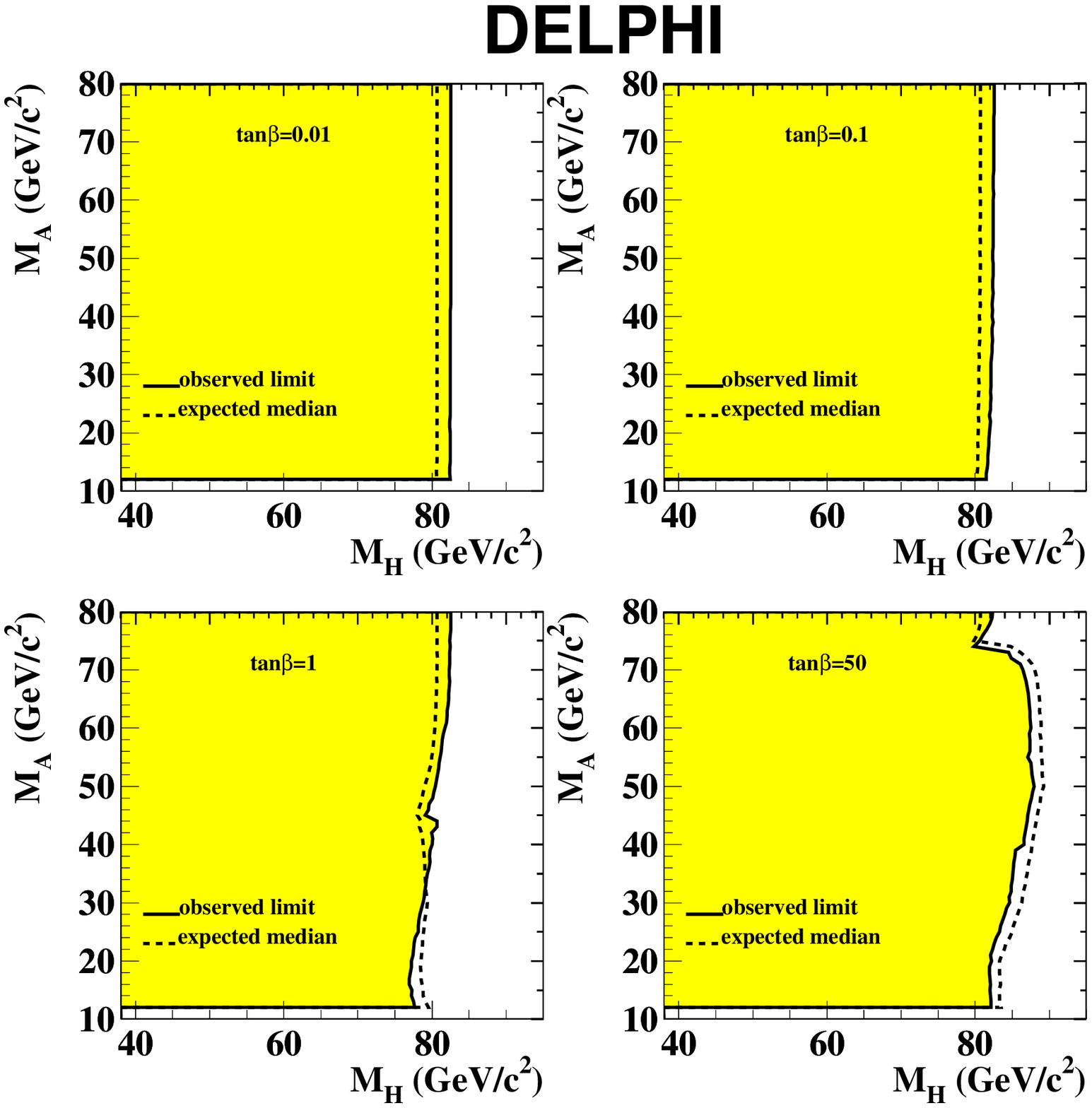,width=15.0cm}
\end{center}
\caption[]{
  The observed and expected exclusion regions at 95\% confidence level in the
  plane of  $\ma$  vs. $\mhp$ in the framework
     of type I Two Higgs Doublet Models. These limits were
  obtained from a combination of the search results in all studied channels, with or without $\wa$ decays,
   at $\sqrt{s}=$ 189--209~GeV, for different values of $\tan\beta$.} 
\label{fig:limitmamh}
\end{figure}

The resulting limits at 95\% confidence level are shown in  Figs.~\ref{fig:limit},~\ref{fig:limitwa} 
and~\ref{fig:limitmamh}
for the two scenarios as functions of the model parameters.
The expected 
lower limits on the mass have been obtained as the
median\footnote{The median is calculated as the value which has 50\% of the limits of the
simulated experiments below it and similarly, the $\pm$ 1$\sigma$ estimations
correspond to 84\% and 16\% of the simulated experiments.}  of a large number of simulated
experiments.  

If the $\wa$ decay is forbidden, a lower H$^{\pm}$
mass limit of $\mhp >$ 74.4~GeV/$c^2$ can be set at 
95\% confidence level, for any branching ratio BR(H
$\rightarrow \tn$). The lower mass limit corresponds to a branching ratio of about 0.3.
The minimum of the expected limits is 76.3~GeV/$c^2$.
The noticeable difference between observed and expected limits is dominated 
by a small unexcluded region (Fig.~\ref{fig:limit})
around BR=0.35 produced by a small excess of data
in that region in the semileptonic channel. However, this region is excluded at 92\% 
confidence level.

Within type I models, a lower limit on the H$^{\pm}$
mass of $\mhp >$ 76.7~GeV/$c^2$ can be set at 
95\% confidence level, for any $\tan\beta$, for $\ma>12$~GeV/$c^2$.
The expected lower limit on the mass for these conditions was 77.1~GeV/$c^2$.
Table~\ref{tab:limitswa} shows the limits obtained for different values of $\ma$
and $\tan\beta$. The lower limit on the mass for a given $\ma$ or a given $\tan\beta$ and the
absolute lower limit are also shown.

\begin{table}[h!]
\begin{center}
\begin{tabular}{c|c|c|c}
\hline
$\ma$ & \multicolumn{1}{|c}{$\tan\beta=0.01$} &
             \multicolumn{1}{|c}{$\tan\beta=50$} &
             \multicolumn{1}{|c}{minimum}   \\ 
\hline
12     &82.4 (80.7) &82.1 (83.5) &77.6 (77.1) \\
30     &82.5 (80.7) &84.6 (86.3) &78.6 (77.6) \\
50     &82.5 (80.7) &88.0 (89.2) &78.9 (78.4) \\
70     &82.5 (80.6) &86.4 (88.0) &80.2 (79.0) \\     
\hline
minimum&82.4 (80.6) &79.8 (79.9) &76.7 (77.1) \\     
\hline
\end{tabular}
\end{center}
\caption[] {Observed limits for the charged Higgs mass in $\gev /c^2$
at 95\% C.L.  for different values of $\ma$ (in GeV/$c^2$) and
$\tan\beta$. The expected median limit is shown in parenthesis. 
The last column and the last row, show the worst case limits for a fixed mass and any $\tan\beta$ or a fixed
 $\tan\beta$ and any mass.
} 

\label{tab:limitswa}

\end{table}

\begin{figure}[hbpt]
\begin{center}
\epsfig{file=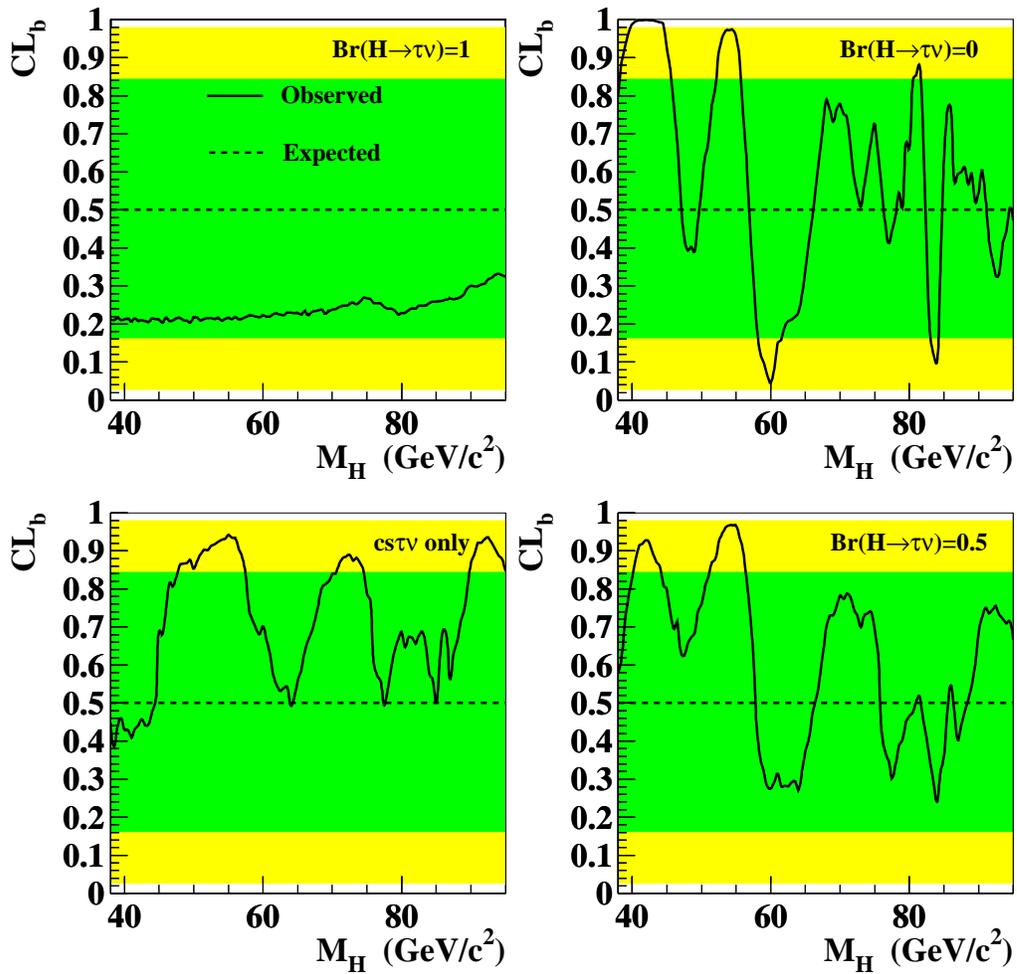,width=15.0cm}
\end{center}
\caption[]{ Confidence level 
for the background-only hypothesis for different branching ratios, under
the assumption that the $\wa$ decay is forbidden. The bottom left figure, shows the $CL_b$ only for the events
selected in the $\HCSTN$.
The full line shows the observed
$CL_b$ and the horizontal dashed line at 0.5 
indicates the expectation in the absence of a signal.
The bands show the one
and two standard deviation regions for this expectation.}
\label{fig:clb}
\end{figure}

\begin{figure}[hbpt]
\begin{center}
\epsfig{file=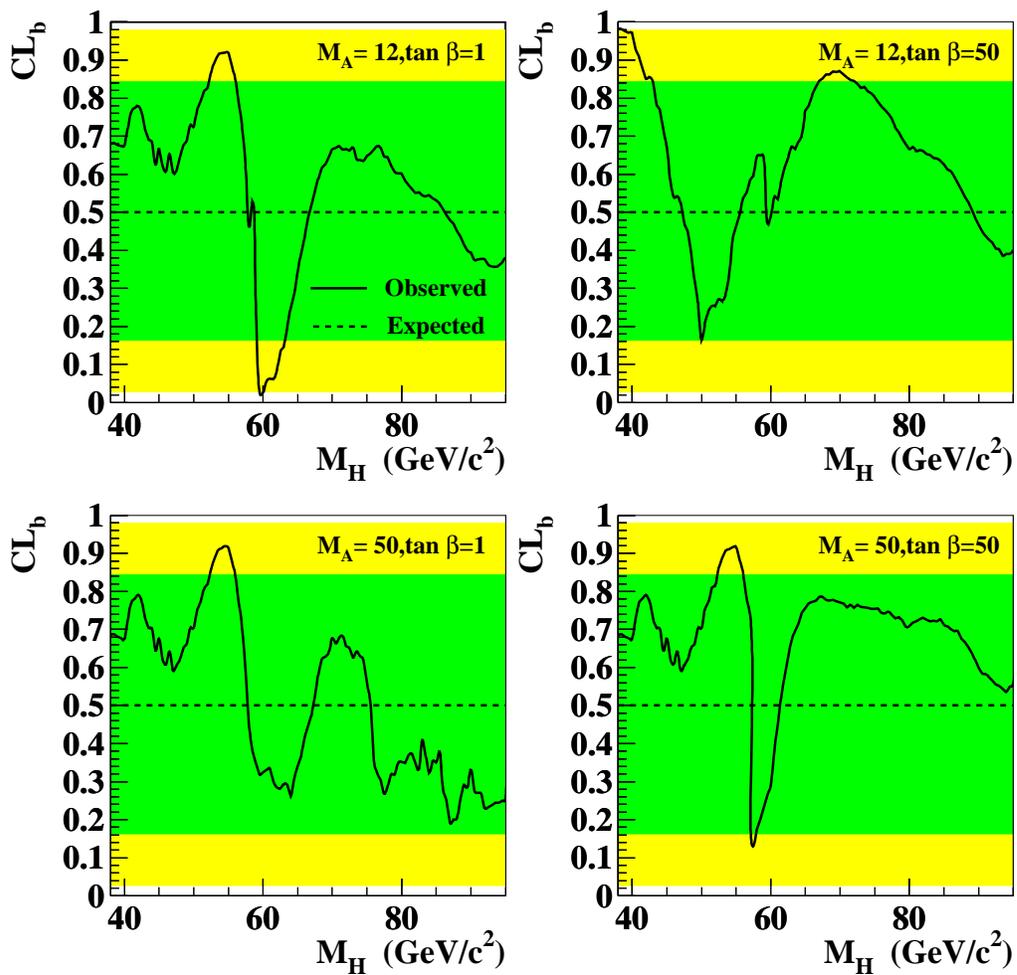,width=15.0cm}
\end{center}
\caption[]{ Confidence level for 
the background-only hypothesis
for different $\tan\beta$ and A
masses. The full line shows the observed
$CL_b$ and the horizontal dashed line at 0.5
indicates the expectation in the absence of a signal.
The bands show the one
and two standard deviation regions for this expectation.}
\label{fig:clbwa}
\end{figure}

Figures~\ref{fig:clb} and \ref{fig:clbwa} show
the observed and expected confidence level for 
the background-only hypothesis\footnote{The confidence level for background-only hypothesis, $CL_b$ is 
defined~\cite{alex,higgs} in such a way that its expectation value is 0.5 in the 
absence of signal. A $CL_b$ value close to 1 indicates a signal-like 
excess of candidates in the data.}. In general a good agreement with this hypothesis is found, 
with
the confidence levels inside the two standard deviation regions. This is true in all cases, except in
a small mass region below 45~GeV/$c^2$ for the $\HCSCS$ decay channel, where the observed confidence level 
corresponds to  3.1
standard deviations. This excess, however, was not found to be compatible with a charged Higgs signal and
therefore considered as a fluctuation for the following reasons. Firstly, the excess is an order of
magnitude smaller than the expected rate from a signal. Secondly, the excess is distributed over much broader mass
range than what would be expected for a charged Higgs signal.
As a consequence, the signal plus background hypothesis
is incompatible with the data with a
confidence level equivalent to more than 13 standard deviations.

\subsection{Cross-section limit}
\begin{figure}[hbpt]
\begin{center}
\epsfig{file=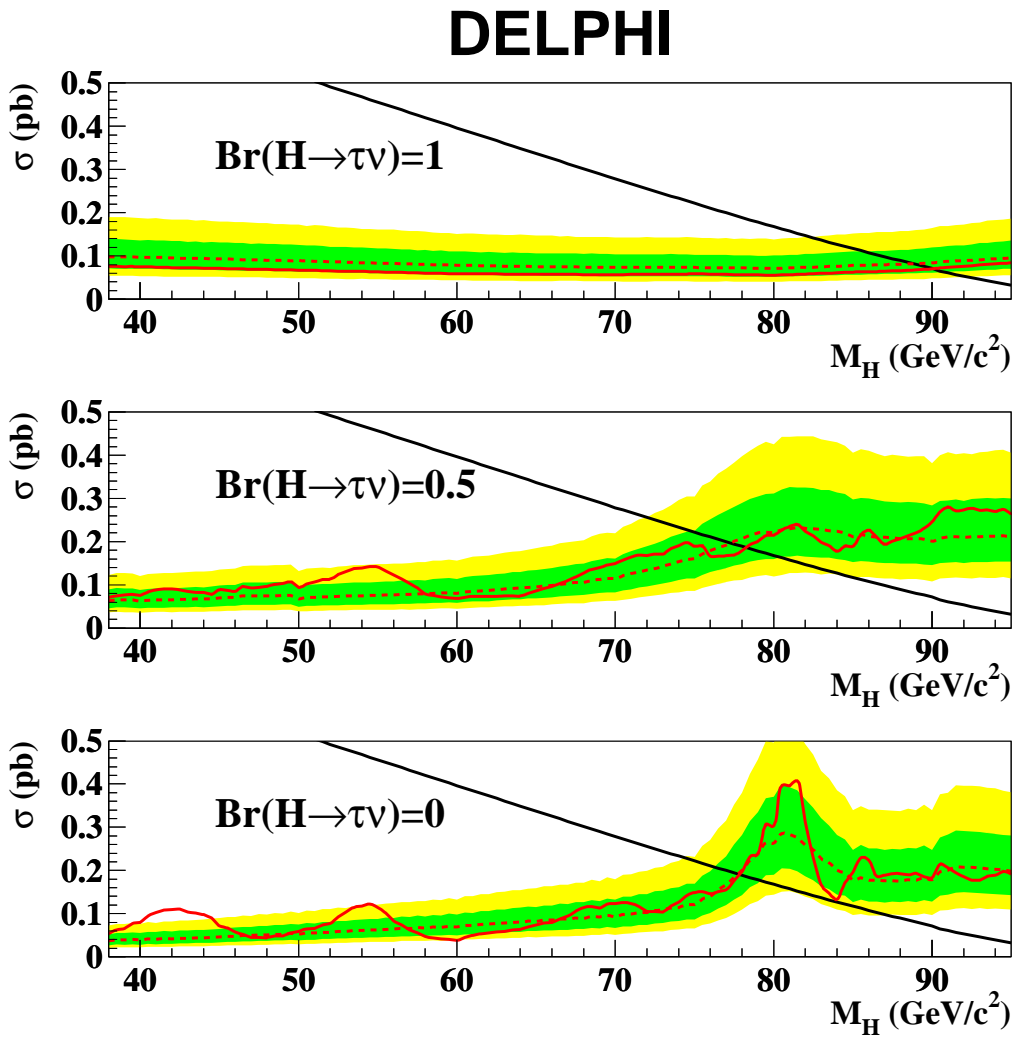,width=15.0cm}
\end{center}
\caption[]{
  Upper limits on the cross-section for charged Higgs boson pair production at
  95\% confidence level,
  for
  different BR($H^- \rightarrow \tn$), under the assumption that the $\wa$ decay
  is forbidden.
  The dashed 
  curve
  shows the expected upper limit with one and two
  standard deviation bands and the solid 
  curve is
  the observed upper limit
  of the cross-section. The solid black diagonal 
  curve
  shows the Two Higgs Doublet Model
  prediction. Cross-sections are given for 206.6 GeV centre-of-mass
  energy.}
\label{fig:xsec}
\end{figure}

\begin{figure}[hbpt]
\begin{center}
\epsfig{file=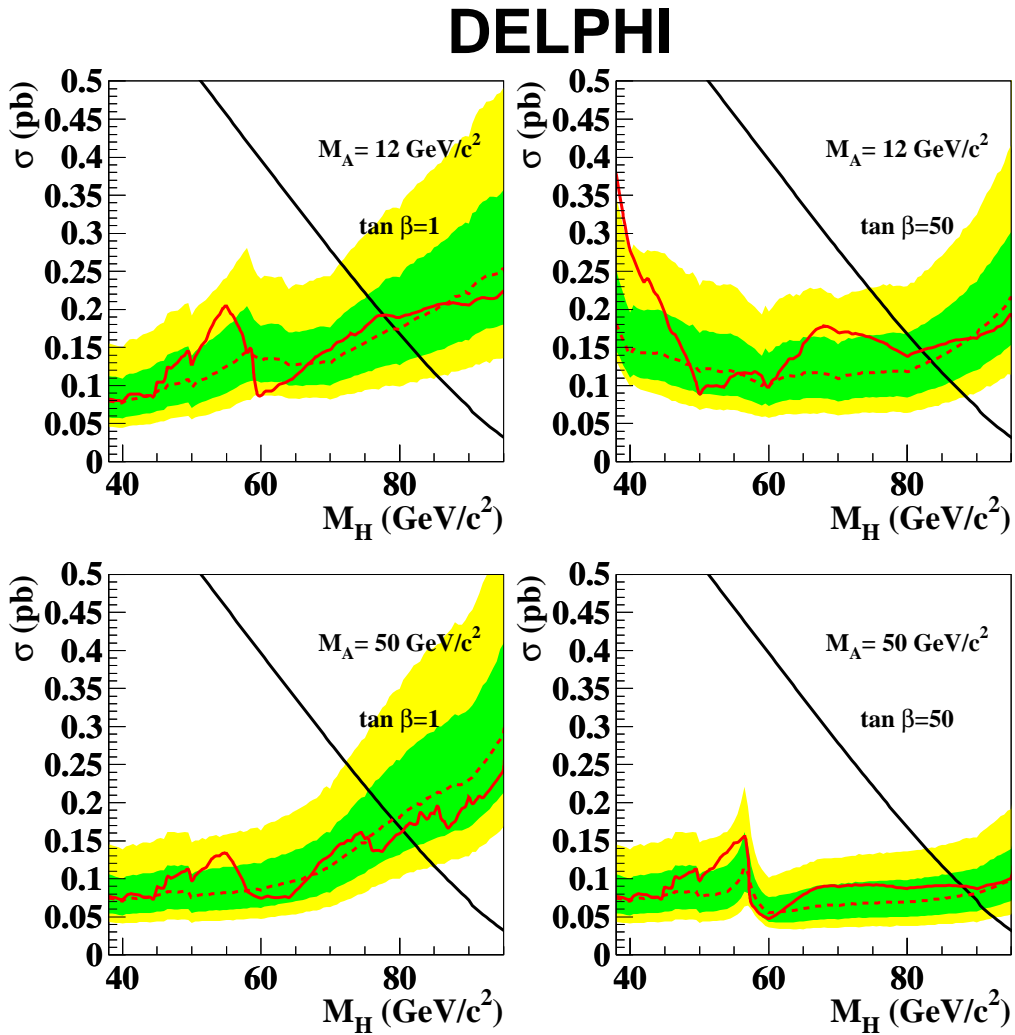,width=15.0cm}
\end{center}
\caption[]{
  Upper limits, at 95\% confidence level, on the production cross-section 
  for a pair of charged Higgs bosons
  as a function of the charged Higgs boson mass, 
  for different $\tan\beta$ and $\ma$ values within type I models.
  The dashed 
  curve
  shows the expected upper limit with one and two
  standard deviation bands and the solid 
  curve  
  the observed upper limit
  of the cross-section. The solid black diagonal 
  curve
  shows the Two Higgs Doublet Model
  prediction. Cross-sections are given for 206.6 GeV centre-of-mass
  energy.}
\label{fig:xsecwa}
\end{figure}

The results are also expressed as 95\% confidence level upper limits
for the charged Higgs boson production cross-section as a function of
the charged Higgs boson mass, for different assumptions on the model 
parameters, i.e. 
leptonic 
branching ratio
for the first scenario and $\ma$ and $\tan\beta$ for the second. 
These cross-section limits were
determined for each mass point by scaling the expected Two Higgs Doublet Model signal
cross-section up or down until the confidence level for signal  hypothesis
reached 95\%. Therefore the only assumption taken from the model is the dependence of the
cross-section 
on
the mass and centre-of-mass energy and thus this approach can be 
considered
model independent to a large extent. Results are summarised in Figures~\ref{fig:xsec}
and~\ref{fig:xsecwa}. 
These cross-sections are given for
206.6 GeV centre-of-mass energy, the maximum energy for which this analysis has a sizable luminosity.

\section{Conclusions}

A search for pair-produced charged Higgs bosons was performed using
the data collected by DELPHI at LEP at centre-of-mass
energies from 189~GeV to 209~GeV 
searching for
the $\HTT$, $\HCSCS$, $\HCSTN$,
$\HWAWA$ and $\HWATN$ 
final states. No significant
excess of candidates over the expected Standard Model background was observed and 
lower limits on the charged
Higgs boson mass were set in two 
scenarios.
Assuming that the 
branching ratio
to $\wa$ is negligible
(type II models or type I with a heavy neutral pseudo-scalar) limits are 
set at 95\% confidence level as a function of the 
branching ratio
to leptons. Results are
shown in Fig.~\ref{fig:limit}. 
The absolute limit is 74.4 GeV$/c^2$ at 95\% confidence level.
Limits were also set within type I models for different neutral pseudo-scalar masses, $\ma>12$~GeV$/c^2$ and 
$\tan\beta$. Results are
shown in Figures~\ref{fig:limitwa} and~\ref{fig:limitmamh}. The absolute limit is 76.7 GeV$/c^2$ at 
95\% confidence level.

To allow a less model-dependent comparison, limits are also expressed in terms of upper bounds
on the cross-section for different sets of the model parameters. Results are shown in 
Figs.~\ref{fig:xsec} and \ref{fig:xsecwa}.

This analysis improves previous searches both by the inclusion of new discriminant techniques and by the
less model-dependent approach allowing more sensitivity and covering a wider range of models and model
parameters. 


\subsection*{Acknowledgments}
\vskip 3 mm
 We are greatly indebted to our technical 
collaborators, to the members of the CERN-SL Division for the excellent 
performance of the LEP collider, and to the funding agencies for their
support in building and operating the DELPHI detector.\\
We acknowledge in particular the support of \\
Austrian Federal Ministry of Science and Traffics, GZ 616.364/2-III/2a/98, \\
FNRS--FWO, Belgium,  \\
FINEP, CNPq, CAPES, FUJB and FAPERJ, Brazil, \\
Czech Ministry of Industry and Trade, GA CR 202/96/0450 and GA AVCR A1010521,\\
Danish Natural Research Council, \\
Commission of the European Communities (DG XII), \\
Direction des Sciences de la Mati$\grave{\mbox{\rm e}}$re, CEA, France, \\
Bundesministerium f$\ddot{\mbox{\rm u}}$r Bildung, Wissenschaft, Forschung 
und Technologie, Germany,\\
General Secretariat for Research and Technology, Greece, \\
National Science Foundation (NWO) and Foundation for Research on Matter (FOM),
The Netherlands, \\
Norwegian Research Council,  \\
State Committee for Scientific Research, Poland, 2P03B06015, 2P03B1116 and
SPUB/P03/178/98, \\
JNICT--Junta Nacional de Investiga\c{c}\~{a}o Cient\'{\i}fica 
e Tecnol$\acute{\mbox{\rm o}}$gica, Portugal, \\
Vedecka grantova agentura MS SR, Slovakia, Nr. 95/5195/134, \\
Ministry of Science and Technology of the Republic of Slovenia, \\
CICYT, Spain, AEN96--1661 and AEN96-1681,  \\
The Swedish Research Council,      \\
Particle Physics and Astronomy Research Council, UK, \\
Department of Energy, USA, DE--FG02--94ER40817. \\

\clearpage

\end{document}